\documentstyle[aps,prb,twocolumn,eqsecnum,epsfig,floats]{revtex}

\begin{document}

\draft
\preprint{}

\twocolumn[\hsize\textwidth\columnwidth\hsize\csname @twocolumnfalse\endcsname

\title{Noise of a Quantum-Dot System in the Cotunneling Regime}

\author{Eugene V. Sukhorukov, Guido Burkard, and Daniel Loss}
\address{Department of Physics and Astronomy,
         University of Basel,\\
         Klingelbergstrasse 82,
         CH--4056 Basel, Switzerland}

\maketitle
\begin{abstract}
We study the noise of the cotunneling current through one or several
tunnel-coupled quantum dots in the Coulomb blockade regime. The various
regimes of weak and strong, elastic and inelastic cotunneling are
analyzed for quantum-dot systems (QDS) with few-level, nearly-degenerate, and
continuous electronic spectra. We find that in contrast to sequential
tunneling where the noise is either Poissonian (due to uncorrelated
tunneling events) or sub-Poissonian (suppressed by charge conservation on
the QDS), the noise in inelastic cotunneling can be super-Poissonian due
to switching between QDS states carrying currents of different strengths.
In the case of weak cotunneling we prove a non-equilibrium
fluctuation-dissipation theorem which leads to a universal expression for
the noise-to-current ratio (Fano factor).
In order to investigate strong cotunneling we develop a microscopic theory
of cotunneling based on the density-operator formalism and using the 
projection operator technique. The master equation for the QDS and the
expressions for current and noise in cotunneling in terms of the stationary
state of  the QDS are derived and applied to QDS with a nearly degenerate and
continuous spectrum.
\end{abstract}


\vskip2pc]
\narrowtext

\section{Introduction}
\label{introduction}

In recent years, there has been great interest in 
transport properties of strongly interacting mesoscopic
systems.\cite{nato} As a rule, the electron interaction effects
become stronger with the reduction of the system size, since the
interacting electrons have a smaller chance to avoid each other. 
Thus it is not surprising that an ultrasmall quantum dot connected 
to leads in the transport regime, being under additional control 
by metallic gates, provides a unique possibility to study strong 
correlation effects both in the leads and in the dot itself.\cite{review1}
This has led to a large number of publications on quantum dots,
which investigate situations where the current acts as a probe of correlation effects.
Historically, the nonequilibrium current fluctuations (shot noise) were 
initially considered as a serious
problem for device applications of quantum dots
\cite{problem1,Korotkov1,Korotkov2} rather than as a
fundamental physical phenomenon.
Later it became clear that 
shot noise is an interesting
phenomenon in itself,\cite{review2} because it
contains additional information about correlations, which is not contained, e.g., in
the linear response conductance and
can be used as a further approach to study transport in quantum dots,
both theoretically
\cite{Korotkov1,Korotkov2,Hershfield1,Halperin,Hanke1,%
Hanke2,Krech1,Krech2,Hung,Anda,Wang,Hershfield2,%
Yamaguchi,Ding,Hershfield3,Korotkov3,LS,Mahn-Soo}
and experimentally.\cite{Birk}

Similarly, the majority of papers on the noise of quantum dots consider
the sequential (single-electron) tunneling regime, where a classical description
(the so-called ``orthodox'' theory) is applicable.\cite{AL}
We are not aware of any discussion in the literature
of the shot noise induced by a cotunneling
(two-electron, or second-order) current,\cite{averinazarov,Glattli} except
Ref.~\onlinecite{LS}, where the particular case of weak cotunneling (see below)
through a double-dot (DD) system is considered. Again, this might be because
until very recently cotunneling has been regarded as a minor contribution
to the sequential tunneling current, which spoils the precision of single-electron
devices due to leakage.\cite{problem2}
However, it is now well understood that cotunneling 
is interesting in itself, since it is responsible for strongly correlated
effects such as the Kondo effect in quantum dots,\cite{Kondo1,Kondo2}
or can be used as a probe of two-electron entanglement and nonlocality,\cite{LS}
etc.

In this paper we present a thorough analysis of the shot noise
in the cotunneling regime.
Since the single-electron ``orthodox'' theory cannot be applied to this
case, we first develop a microscopic theory of cotunneling 
suitable for the calculation of the shot noise in
Secs.~\ref{FDT} and \ref{microscopic}.
[For an earlier microscopic theory of transport
through quantum dots
see Refs.~\onlinecite{Schoeller1,Schoeller2,Schoeller3}.] 
We consider the transport through a quantum-dot system (QDS)
in the Coulomb blockade (CB) regime, in which the quantization of
charge on the QDS leads to a suppression of the sequential tunneling
current except under certain resonant conditions.  We consider the transport
away from these resonances and study the next-order contribution to
the current, the so-called cotunneling current.\cite{averinazarov,Glattli}
In general, the QDS can contain several dots, which can be 
coupled by tunnel junctions, the double dot (DD) being a particular
example.\cite{LS}
The QDS is assumed to be weakly coupled to external metallic leads which are
kept at equilibrium with their associated reservoirs at the chemical
potentials $\mu_l$, $l=1,2$,  where the  currents $I_l$ can be measured
and the average current $I$ 
through the QDS is defined by Eq.~(\ref{current-noise}). 

\begin{figure}
  \begin{center}
    \leavevmode
\epsfxsize=8cm
\epsffile{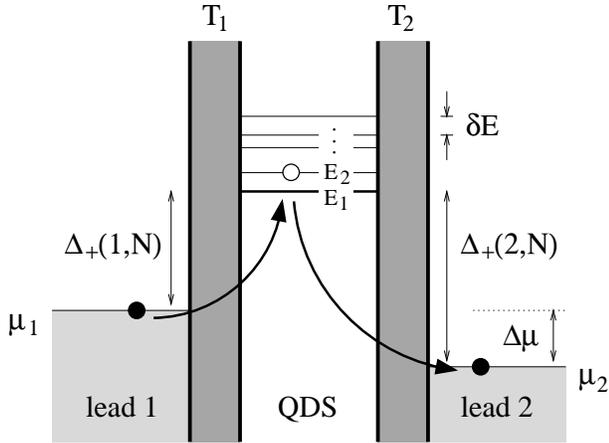}
  \end{center}
\caption{
Schematic representation of the quantum dot system (QDS) coupled to
two external leads $1$ and $2$ (light grey) via tunneling barriers
(dark grey), where the energy scale is drawn vertically.
The tunneling between the QDS and the leads
$l=1,2$ is parametrized by the tunneling amplitudes $T_l$,
where the lead and QDS quantum numbers $k$ and $p$ have
been dropped for simplicity, see Eq.~(\ref{tunneling}).
The leads are at the chemical potentials $\mu_{1,2}$, with an applied
bias $\Delta\mu=\mu_1 -\mu_2$.
The (many-particle) eigenstates of the QDS with one
added electron ($N+1$ electrons in total) are indicated by their energies
$E_1$, $E_2$, etc., with average level-spacing $\delta E$.
The energy cost for adding a particle from the Fermi level of lead $l$
to the N-electron QDS is denoted by $\Delta_+(l,N)>0$ and is strictly
positive in the CB regime.
Note that the energies $\Delta_-(l,N)$ for removing particles from the QDS
containing $N$ electrons are positive as well, and are not drawn here.
The cotunneling process is visualized by two arrows, leading from the
initial state in, say, lead 1 (full circle), via a virtual state on the
QDS (open circle), to the final state in lead 2 (full circle).
}
\label{energies}
\end{figure}

Before proceeding with our analysis 
we briefly review the results available in the literature 
on noise of sequential tunneling. For doing this, we introduce
right from the beginning all relevant physical 
parameters, namely the bath temperature $T$, bias 
$\Delta\mu = \mu_1-\mu_2$, charging energy $E_C$,
average level spacing $\delta E$, and the level width 
$\Gamma=\Gamma_1+\Gamma_2$ of the QDS, where the tunneling rates
$\Gamma_{l}=\pi \nu |T_l|^2$
to the leads $l=1,2$ are expressed in terms of tunneling amplitudes $T_l$
and the density of states $\nu$ evaluated at the Fermi energy of the leads.
In Fig.~\ref{energies} the most important parameters are shown schematically.
This variety of parameters shows
that many different regimes of the CB  are possible. 
In the linear response regime, $\Delta\mu\ll k_BT$,
the thermal noise \cite{thermal} is given by the equilibrium
fluctuation-dissipation theorem (FDT).\cite{FDT} 
Although the cross-over
from the thermal to nonequilibrium noise is of our interest
(see Sec.~\ref{FDT}),
in this section we discuss the shot noise alone and set $T=0$.
Then the noise at zero frequency $\omega=0$, when $\delta I_2=-\delta I_1$, 
can be characterized by one single parameter,
the dimensionless Fano factor $F=S(0)/e|I|$, where 
the spectral density of the noise $S(0)\equiv S_{22}(0)$ is
defined by Eq.~(\ref{current-noise}). The Fano factor acquires
the value $F=1$ for uncorrelated Poissonian noise.

Next we discuss the different CB regimes.
(1) In the limit of large bias $\Delta\mu\gg E_C$, when the CB is suppressed,
the QDS can be viewed as being composed of two tunnel junctions
in series, with the total conductance $G=G_1G_2/(G_1+G_2)$,
where $G_l=\pi e^2\nu\nu_D|T_l|^2$ is the conductance of the tunnel junctions to
lead $l$, and $\nu_D$ is the density of dot states.
Then the Fano factor is given by $F=(G_1^2+G_2^2)/(G_1+G_2)^2$,  
as it has been found in Refs.~\onlinecite{Korotkov1,Korotkov2,Hershfield1}.
Thus, the shot noise is suppressed, $F<1$, and reaches its minimum 
value for the symmetric QDS, $G_1=G_2$, where $F=1/2$.
(2) The low bias regime, $\delta E \ll \Delta\mu\ll E_C$. 
The first inequality $\delta E\ll \Delta\mu$ allows to assume
a continuous spectrum on of the QDS and guarantees that 
the single-electron ``orthodox'' theory based on a classical master equation
can be applied. The second inequality $\Delta\mu\ll E_C$ means that the QDS
is in the CB regime, where the energy cost 
$\Delta_{\pm}(l,N)=E(N\pm 1)-E(N)\mp\mu_l$  
for the electron tunneling from the Fermi level of the lead $l$ 
to the QDS ($+$) and vice versa ($-$)
oscillates as a function of gate voltage between its minimum value $\Delta_{\pm}<0$
(where the energy deficit turns into a gain, $|\Delta_{\pm}|\sim\Delta\mu$)
and its maximum value $\Delta_{\pm}\sim E_C$.
Here, $E(N)$ denotes the ground-state energy of the $N$-electron QDS.
Thus the current $I$ as a function of the gate
voltage consists of the CB peaks which are at the degeneracy points $\Delta_{\pm}<0$,
where the number of electrons on the QDS fluctuates between $N$ and $N+1$
due to single-electron tunneling. The peaks are separated by plateaus,
where the single-electron tunneling is blocked because of the finite energy cost
$\Delta_{\pm}>0$ and thus the sequential tunneling current vanishes. At the peaks the 
current is given by $I=e\gamma_1\gamma_2/(\gamma_1+\gamma_2)$,
while the Fano factor has been reported
\cite{Korotkov2,Hershfield1,Halperin,Hanke1,Hanke2} 
to be equal to $F=(\gamma_1^2+\gamma_2^2)/(\gamma_1+\gamma_2)^2$, 
$1/2<F<1$, where $\gamma_1=e^{-2}G_1|\Delta_{+}(1,N)|$ and 
$\gamma_2=e^{-2}G_2|\Delta_{-}(2,N+1)|$ 
are the tunneling rates to the QDS from lead 1
and from the QDS to lead 2, respectively. Within the ``orthodox'' theory
tunneling is still possible between the peaks
at finite temperature due to thermal activation
processes, and then the Fano factor approaches the Poissonian value $F=1$
from below. (3) Finally, the limit $\Gamma\ll \Delta\mu\ll \delta E$
is similar to the previous case, with the only difference that the dot spectrum is discrete.
The sequential tunneling picture can still be applied; the result for the 
Fano factor at the current peak is
$F=(\Gamma_1^2+\Gamma_2^2)/(\Gamma_1+\Gamma_2)^2$,
so that again $1/2<F<1$.\cite{Hershfield2}

We would like to emphasize the striking similarity of the Fano factors in all three 
regimes, where they also resemble the Fano factor of the noninteracting double-barrier system.\cite{review2}
The Fano factors in the first and second regimes 
become even equal if the ground-state level of the QDS
lies exactly in the middle between the Fermi levels of lead 1 and 2,
$|\Delta_{+}|=|\Delta_{-}|$.
We believe that this ``ubiquitous'' \cite{Hershfield1} double-barrier character of 
the Fano factor can be interpreted 
as being the result of the natural correlations imposed by charge conservation
rather than by interaction effects.
Indeed, in the transport through a double-barrier
tunnel junction each barrier can be thought of as an independent source 
of Poissonian noise. And although in the second regime the CB is explicitly 
taken into account, the stronger requirement of charge conservation
at zero frequency, $\delta I_1+\delta I_2=0$, has to be satisfied, which leads
to additional correlations between the two sources of noise and to 
a suppression of the noise below the Poissonian value.
At finite frequency (but still in the classical range defined as $\omega\ll \Delta\mu, E_C$) 
temporary charge accumulation on the QDS is allowed, and for frequencies larger
than the tunneling rate, 
$\omega\gg\gamma_{1,2}$, the conservation of charge does not need to
be satisfied, while the noise power $S_{22}$ approaches its 
Poissonian value from below, and the cross correlations vanish, $S_{12}=0$.
\cite{comment1} 
Based on this observation we expect that the direct measurement of 
interaction effects in noise is only possible either in the quantum
(coherent) CB regime \cite{Hershfield2} $\Delta\mu\sim\Gamma$ 
or in the Kondo regime,\cite{Yamaguchi,Ding,Hershfield3}
where both charge conservation and many-electron effects
lead to a suppression of the noise.
Another example is the noise in the quantum regime, 
$\Delta\mu\leq \omega\sim E_C$,
where it contains singularities associated with the ``photo-assisted transitions'' 
above the Coulomb gap $\Delta_{\pm}$.
\cite{Korotkov3,LS,Josephson-frequency}

To conclude our brief review we would like to emphasize again that while
the zero-frequency shot noise in the sequential tunneling regime is
always suppressed below its full Poissonian value as a result of charge
conservation (interactions suppressing it further), we find that,
in the present work the shot noise in the
cotunneling regime\cite{cotunneling1}  
is either Poissonian $F=1$  (elastic or weak inelastic cotunneling)
or, rather surprisingly, non-Poissonian $F\neq 1$ 
(strong inelastic cotunneling). 
Therefore the non-Poissonian noise in QDS can be considered as being a fingerprint
of inelastic cotunneling. This difference of course
stems from the different physical origin of the noise in the cotunneling regime,
which we discuss next.
Away from the sequential tunneling peaks, $\Delta_{\pm}>0$,
single-electron tunneling is blocked, and 
the only elementary tunneling process which is compatible with
energy conservation is the simultaneous tunneling of two electrons
called cotunneling\cite{averinazarov,Glattli}.
In this process one electron tunnels, say, from lead $1$
into the QDS, and the other
electron tunnels from the QDS into lead $2$ 
with a time delay on the order of $\Delta_{\pm}^{-1}$ (see Ref.\ \onlinecite{LS}).
This means that in the range of frequencies, 
$\omega\ll\Delta_{\pm}$,
(which we assume in our paper) 
the charge on the QDS does not fluctuate,
and thus in contrast to the sequential tunneling 
the correlation imposed by charge conservation is not relevant
for cotunneling. Furthermore, 
in the case of elastic cotunneling ($\Delta\mu<\delta E$), 
where the state of the QDS remains 
unchanged, the QDS can be effectively regarded 
as a single barrier. Therefore, subsequent elastic cotunneling events
are uncorrelated, and the noise is Poissonian with $F=1$.
On the other hand, this is not so for inelastic cotunneling($\Delta\mu>\delta E$), 
where the internal state
of the QDS is changed, thereby changing the conditions for the subsequent 
cotunneling event. Thus, in this case the QDS switches between 
different current states, and this creates a correction to noise $\Delta S$, 
so that the total noise is non-Poissonian,
and can become super-Poissonian. 
The other mechanism
underlying super-Poissonian noise is the excitation of high
energy levels (heating) of the QDS
caused by multiple inelastic cotunneling transitions and leading to
the additional noise $\Delta S_h$. Thus the total noise can be written as 
$S=eI+\Delta S_h+\Delta S$.
For other cases exhibiting super-Poissonian noise
(in the strongly non-linear bias regime) see Ref.~\onlinecite{review2}.

According to this picture we consider the following  different regimes
of the inelastic cotunneling.
We first discuss the {\em weak cotunneling} regime $w\ll w_{\rm in}$, where
$w\sim\Gamma_1\Gamma_2\Delta\mu/\Delta_{\pm}^2$ 
is the average rate of the inelastic cotunneling
transitions on the QDS [see Eqs.~(\ref{master-6}-\ref{golden-rule})], 
and $w_{\rm in}$ is the intrinsic relaxation rate of the QDS to its equilibrium 
state due to the coupling to the environment.
In this regime the
cotunneling happens so rarely that the QDS always relaxes to its
equilibrium state before the next electron passes through it. Thus we
expect no correlations between cotunneling events in
this regime, and the zero-frequency noise  is going to take on its
Poissonian value with Fano factor $F=1$, as first obtained for a
special case in Ref.~\onlinecite{LS}. This result is
generalized in Sec.~\ref{FDT}, where we find a universal relation
between noise and current of single-barrier tunnel
junctions and, more generally, of the QDS 
in the first nonvanishing order in the tunneling
perturbation $V$. Because of the universal
character of the results Eqs.~(\ref{SB-FDT}) and (\ref{DB-FDT})
we call them the nonequilibrium FDT in analogy with linear
response theory.

Next, we consider {\em strong cotunneling}, i.e.\ $w\gg w_{\rm in}$.
The microscopic theory of the transport and noise in this
regime based on a projector operator technique is developed in
Sec.~\ref{microscopic}.
In the case of a {\em few-level} QDS, $\delta E\sim E_C$,
\cite{comment2}
noise turns out to be non-Poissonian, as we have discussed above,
and this effect can be estimated as follows.
The QDS is switching between 
states with the different currents $I\sim ew$, and we find 
$\delta I\sim ew$.  The QDS stays in each state for the time 
$\tau\sim w^{-1}$. Therefore, for the positive correction to
the noise power we get $\Delta S\sim\delta I^2\tau\sim e^2w$,
and the estimate for the correction to the Fano factor
follows as $\Delta S/eI\sim 1$. A similar result is
expected for the noise induced by heating, $\Delta S_h$, which can roughly be
estimated by assuming an equilibrium distribution
on the QDS with the temperature $k_BT\sim\Delta\mu$
and considering the additional noise as being thermal,\cite{thermal}
$\Delta S_h\sim G k_BT\sim (eI/\Delta\mu)k_BT\sim eI$.
The characteristic frequency of the noise correction
$\Delta S$ is $\omega\sim w$, with $\Delta S$ vanishing
for $\omega\gg w$ 
(but still in the classical range, $\omega\ll\Delta\mu$). 
In contrast to this, the additional noise 
due to heating, $\Delta S_h$, does not depend on the frequency.

In Sec.~\ref{degenerate} we consider the particular case
of nearly degenerate dot states, in which only few levels
with an energy distance
smaller than $\delta E$ participate in transport, 
and thus heating on the QDS can be neglected. 
Specifically, for a two-level QDS we predict giant (divergent) 
super-Poissonian noise if the off-diagonal transition rates
vanish. The QDS goes into an unstable mode where it switches
between states 1 and 2 with (generally) different currents.
We consider the transport through a double-dot (DD) system 
as an example to illustrate this effect [see Eq.~(\ref{DD-noise})
and Fig.~\ref{noise-corr}].

Finally, we discuss the case of a {\em multi-level} QDS, 
$\delta E\ll E_C$.
In this case the correlations in the cotunneling current
described above do not play an essential role.  
In the regime of low bias, $\Delta\mu\ll (\delta E\, E_C)^{1/2}$,
elastic cotunneling dominates transport,\cite{averinazarov,comment3}
and thus the noise is Poissonian.
In the opposite case of large bias, 
$\Delta_{\pm}\gg \Delta\mu\gg (\delta E\, E_C)^{1/2}$,
the transport is governed by inelastic cotunneling,
and in Sec.~\ref{continuum} we study heating effects which
are relevant in this regime.
For this we use the results of
Sec.~\ref{microscopic} and derive a kinetic
equation for the distribution function $f(\varepsilon)$.
We find three universal regimes 
where $I\sim\Delta\mu^3$, and the Fano factor does not
depend on bias the $\Delta\mu$.
The first is the regime of weak cotunneling, $\tau_{\rm in}\ll\tau_c$,
where $\tau_{\rm in}$ and $\tau_c$ are time scales characterizing
the single-particle dynamics of the QDS.
The energy relaxation time $\tau_{\rm in}$ describes 
the strength of the coupling to the environment
while $\tau_c\sim e\nu_D\Delta\mu/I$
is the cotunneling transition time.
Then  we obtain for the distribution
$f(\varepsilon)=\theta(-\varepsilon)$, 
reproducing the result of Ref.~\onlinecite{averinazarov}.
We also find that $F=1$, in agreement with the FDT proven in
Sec.~\ref{FDT}. The other two regimes 
of strong cotunneling $\tau_{\rm in}\gg\tau_c$ are
determined by the electron-electron scattering time $\tau_{ee}$.
For the cold-electron regime, $\tau_c\ll\tau_{ee}$,
we find the distribution function by solving the integral 
equations (\ref{kineq1}) and (\ref{kernel}), 
while for hot electrons, $\tau_c\gg\tau_{ee}$,
$f$ is given by the Fermi distribution function with
an electron temperature obtained from the energy balance 
equation (\ref{kineq3}). 
We use $f(\varepsilon)$ to calculate the Fano factor,
which turns out to be very close to 1. On the 
other hand, the current depends 
not only on $G_1G_2$ but also on the ratio, $G_1/G_2$,
depending on the cotunneling regime [see Fig.~\ref{cold-hot}].
Details of the calculations are deferred to four appendices.

\section{Model system}

The quantum-dot system (QDS) under study is weakly coupled to
two external metallic leads which are kept
in equilibrium with their associated reservoirs at the chemical
potentials $\mu_l$, $l=1,2$, where the currents $I_l$ can be
measured.
Using a standard tunneling Hamiltonian approach,\cite{Mahan}
we write
\begin{eqnarray}
&& H=H_0+V\,,\quad H_0=H_L+H_S+H_{\rm int}\,, \label{Hamiltonian}   \\
&& H_L=\sum_{l=1,2}\sum_k\varepsilon_{k}c_{lk}^{\dag}c_{lk}\,, \quad
H_S=\sum_p\varepsilon_pd_p^{\dag}d_p\,, \label{QDS-leads} \\
&& V=\sum_{l=1,2}(D_l+D^{\dag}_l),\quad
D_l=\sum_{k,p}T_{lkp}c_{lk}^{\dag}d_p\,,\label{tunneling}
\end{eqnarray}
where  the terms $H_L$ and $H_S$  describe
the leads and QDS, respectively (with $k$ and $p$
from a complete set of quantum numbers),
and tunneling between leads  and  QDS is described 
by the perturbation $V$.
The interaction term $H_{\rm int}$ is specified below.
The $N$-electron QDS is in the cotunneling regime where there is  
a finite energy cost 
$\Delta_{\pm}(l,N)>0$  
for the electron tunneling from the Fermi level of the lead $l$ 
to the QDS ($+$) and vice versa ($-$),
so that only processes of second order in $V$ are allowed.

To describe the transport through the QDS we apply standard methods\cite{Mahan}
and adiabatically switch on the perturbation $V$ in the distant past,
$t=t_0\to - \infty$. The perturbed state of the system
is described by the time-dependent density matrix
$\rho(t)=e^{-iH(t-t_0)}\rho_0 e^{iH(t-t_0)}$,
which can be written as
\begin{equation}
\rho(t)=e^{-iL(t-t_0)}\rho_0\,,\quad
LA\equiv\left[H,A\right]\,,\quad \forall A\,,
\label{dmatrix}
\end{equation}
with the help of the Liouville operator $L=L_0+L_V$.\cite{FS} 
Here $\rho_0$ is the grand canonical density matrix of the
unperturbed system,
\begin{equation}
\rho_0=Z^{-1}e^{-K/k_BT},
\label{rho0}
\end{equation}
where we set $K=H_0-\sum_l\mu_lN_l$.

Because of tunneling the total number of electrons in each lead
$N_l=\sum_kc_{lk}^{\dag}c_{lk}$ is no longer conserved.
For the outgoing currents $\hat I_l=e\dot N_l$ we have
\begin{equation}
\hat I_l=ei\left[V,N_l\right] =ei(D^{\dag}_l-D_l)\,.
\label{currents}
\end{equation}
The observables of interest are the average current $I\equiv I_2=-I_1$
through the QDS,
and the spectral density of the noise 
$S_{ll'}(\omega)=\int dt S_{ll'}(t)\exp(i\omega t)$,
\begin{equation}
I_l={\rm Tr}\rho(0)\hat I_l,\quad
S_{ll'}(t)={\rm Re}\,{\rm Tr}\,\rho(0)\delta I_l(t)\delta I_{l'}(0)\,,
\label{current-noise}
\end{equation}
where $\delta I_l=\hat I_l-I_l$.
Below we will use the interaction representation
where Eq.~(\ref{current-noise}) can be rewritten by replacing
$\rho(0)\to\rho_0$ and $\hat I_l(t)\to U^{\dag}(t)\hat I_l(t)U(t)$, with
\begin{equation}
U(t)=T\exp\left[-i\int^{t}_{-\infty}dt'\,V(t')\right]\,.
\label{U-Operator}
\end{equation}
In this representation, the time dependence of all operators is governed by the
unperturbed Hamiltonian $H_0$.

\section{Non-equilibrium fluctuation-dissipation theorem
for tunnel junctions}
\label{FDT}

In this section we prove the universality of noise of tunnel junctions
in the weak cotunneling regime $w\ll w_{\rm in}$ keeping the first
nonvanishing order in the tunneling Hamiltonian $V$.
Since our final results Eqs.~(\ref{SB-FDT}), (\ref{conductance2}),
(\ref{SB-FDT2}), and (\ref{DB-FDT}) can be applied to quite general
systems out-of-equilibrium
we call this result the non-equilibrium fluctuation-dissipation theorem (FDT).
In particular, the geometry of the QDS and the interaction $H_{\rm int}$ are 
completely arbitrary for the discussion of 
the non-equilibrium FDT in this section.
Such a non-equilibrium FDT was derived for single barrier junctions
long ago.\cite{Rogovin}
We will need to briefly review this case which allows us then to generalize
the FDT to QDS considered here in the most direct way.

\subsection{Single-barrier junction}
\label{FDT-Single}

The total Hamiltonian of the junction
[given by Eqs.~(\ref{Hamiltonian})-(\ref{tunneling})]
and the currents Eq.~(\ref{currents}) have to be replaced by
$H=H_L+H_{\rm int}+V$, where
\begin{eqnarray}
&& V=A+A^{\dag}\,,\quad
A=\sum_{k,k'}T_{kk'}c_{2k}^{\dag}c_{1k'}\,,
\label{SB-tunneling} \\
&& \hat I_2=-\hat I_1=ei\left[V, N_2\right]=ei\left(A^{\dag}-A\right)\,.
\label{SB-current}
\end{eqnarray}
For the sake of generality,
we do not specify the interaction $H_{\rm int}$ in this section,
nor the electron spectrum
in the leads, and the geometry of our system.

Applying the standard interaction representation technique,\cite{Mahan}
we expand the expression (\ref{U-Operator}) for $U(t)$ and
keep only first non-vanishing contributions in $V$, obtaining
\begin{equation}
I(t)=i\int\limits^{t}_{-\infty}dt'\langle\left[V(t'),\hat I_2(t)\right]\rangle\,,
\label{SB-current2}
\end{equation}
where we use the notation $\langle\ldots\rangle={\rm Tr}\rho_0 (\ldots )$.
Analogously, we find that the first non-vanishing
contribution to the noise power $S(\omega)\equiv S_{22}(\omega)$
is given by
\begin{equation}
S(\omega)=\frac{1}{2}\int\limits^{\infty}_{-\infty}dt\,
e^{i\omega t}\langle\{\hat I_2(t), \hat I_2(0)\}\rangle\,,
\label{SB-power}
\end{equation}
where $\{\ldots\}$ stands for anticommutator, 
and $I_2^2=0$ in leading order.

We notice that in Eqs. (\ref{SB-current2}) and (\ref{SB-power})
the terms $\langle AA\rangle$ and $\langle A^{\dag}A^{\dag}\rangle$
are responsible for Cooper pair tunneling and vanish in the case
of normal (interacting) leads. Taking this into account
and using Eqs. (\ref{SB-tunneling}) and (\ref{SB-current})
we obtain
\begin{eqnarray}
&&  I=e\int\limits^{\infty}_{-\infty}dt\,
\langle\left[A^{\dag}(t),A(0)\right]\rangle\,,
\label{SB-current3} \\
&& S(\omega)=e^2\int\limits^{\infty}_{-\infty}dt\,
\cos (\omega t)\langle\{A^{\dag}(t),A(0)\}\rangle\,,
\label{SB-noise1}
\end{eqnarray}
where we also used
$\langle A^{\dag}(t)A(0)\rangle=\langle A^{\dag}(0)A(-t)\rangle$.

Next we apply the spectral decomposition
to the correlators Eqs.~(\ref{SB-current3}) and (\ref{SB-noise1}), 
a similar procedure to that which also leads
to the equilibrium fluctuation-dissipation theorem.
The crucial observation is that $[H_0,N_l]=0$, $l=1,2$ (we stress that
it is only the tunneling Hamiltonian $V$ which does not commute with $N_l$, while
all interactions do not change the number of electrons in the leads).
Therefore, we are allowed to use for our spectral decomposition the basis  
$|{\bf n}\rangle=| E_{{\bf n}},N_1,N_2\rangle$ of eigenstates
of the operator $K=H_0-\sum_l\mu_l N_l$, which also diagonalizes the grand-canonical
density matrix $\rho_0$
[given by Eq.~(\ref{rho0})], $\rho_{{\bf n}}=\langle {\bf n}|\rho_0 |{\bf n} \rangle
=Z^{-1}\exp[-E_{{\bf n}}/k_BT]$. 
Next we introduce the spectral function,
\begin{eqnarray}
&&{\cal A}(\omega) =
2\pi \sum_{ {\bf n}, {\bf m}}(\rho_{{\bf n}} +\rho_{{\bf m}})
|\langle {\bf m}|A|{\bf n}\rangle|^2\nonumber\\
&&\qquad\qquad\qquad\times\delta(\omega+E_{{\bf n}}-E_{{\bf m}})\,,
\label{spectral}
\end{eqnarray}
and rewrite Eqs.~(\ref{SB-current3}) and (\ref{SB-noise1})
in the matrix form in the basis $| {\bf n}\rangle$  
taking into account that 
the operator $A$ creates (annihilates) an electron in the lead 2 (1)
[see Eq.~(\ref{SB-tunneling})].
We obtain following expressions
\begin{eqnarray}
&&  I(\Delta\mu)=e\tanh\left[\frac{\Delta\mu}{2k_BT}\right]{\cal A}(\Delta\mu)\,,
\label{SB-current4} \\
&& S(\omega,\Delta\mu)=\frac{e^2}{2}\sum_{\pm}
{\cal A}(\Delta\mu\pm\omega)\,,
\label{SB-noise2}
\end{eqnarray}
where $\Delta\mu=\mu_1-\mu_2$. From these equations our main result follows
\begin{equation}
S(\omega,\Delta\mu)=\frac{e}{2}\sum_{\pm}
\coth\left[\frac{\Delta\mu\pm\omega}{2k_BT}\right]I(\Delta\mu\pm\omega)\,,
\label{SB-FDT}
\end{equation}
where we have neglected contributions of order $\Delta\mu/\varepsilon_F,\omega/\varepsilon_F\ll 1$.
We call the relation (\ref{SB-FDT}) non-equilibrium fluctuation-dissipation theorem 
because of its general validity
(we recall that no assumptions on geometry or interactions were made).

The fact that the spectral function Eq.~(\ref{spectral})  depends
only on one parameter can be used to obtain further useful relations.
Suppose that in addition to the bias $\Delta\mu$
a small perturbation of the form $\delta\mu e^{-i\omega t}$ is applied to the 
junction. This perturbation generates an ac current $\delta I(\omega,\Delta\mu)e^{-i\omega t}$
through the barrier, which depends on both parameters, $\omega$ and $\Delta\mu$.
The quantity of interest is the linear response conductance
$G(\omega, \Delta\mu)=e\delta I(\omega,\Delta\mu)/\delta\mu$.
The perturbation $\delta\mu$ can be taken into account in a standard way by multiplying 
the tunneling amplitude $A(t)$ by a phase factor $e^{-i\phi(t)}$, where 
$\dot\phi=\delta\mu e^{-i\omega t}$. Substituting the new amplitude 
into Eq.~(\ref{SB-current2}) and expanding the current with respect to
$\delta\mu$, we arrive at the following result,
\begin{equation}
{\rm Re}\, G(\omega,\Delta\mu)=
\frac{ie^2}{\omega}\int\limits^{\infty}_{-\infty}
dt\sin(\omega t)\langle [A^{\dag}(t),A(0)]\rangle\,.
\label{conductance1}
\end{equation}
Finally, applying the spectral decomposition to this equation we obtain  
\begin{equation}
(2/e)\omega\,{\rm Re}\,G(\omega,\Delta\mu)=
I(\Delta\mu+\omega)-I(\Delta\mu-\omega),
\label{conductance2}
\end{equation}
which holds for a general nonlinear $I$ vs $\Delta\mu$
dependence. From this equation and from Eq.~(\ref{SB-FDT}) it
follows that the noise power at zero frequency can be expressed
through the conductance at finite frequency as follows
\begin{eqnarray}
&&S(0,\Delta\mu)+S(0,-\Delta\mu)=\nonumber\\
&&\quad\quad\quad\quad 2\omega\coth\left[\frac{\omega}{2k_BT}\right]
{\rm Re}\,G(\omega,0)|_{\omega\to\Delta\mu}.
\label{SB-FDT2}
\end{eqnarray}
And for the noise power at zero bias we obtain 
$S(\omega, 0)=\omega\coth(\omega/2k_BT){\rm Re}\,G(\omega,0)$, which is 
the standard equilibrium FDT.\cite{FDT}
Eq.~(\ref{SB-FDT}) reproduces the result
of Ref.~\onlinecite{Rogovin}.  
The current is not necessary linear
in $\Delta\mu$ (the case of tunneling into a Luttinger liquid
\cite{KaneFisher} is an obvious example),
and in the limit $T,\omega\to 0$ we find the Poissonian
noise, $S=eI$. In the limit $T,\Delta\mu\to 0$, the 
quantum noise becomes $S(\omega)=e[I(\omega)-I(-\omega)]/2$. 
If $I(-\Delta\mu)=-I(\Delta\mu)$, we get $S(\omega)=eI(\omega)$,
and thus $S(\omega)$ can be obtained from $I(\Delta\mu\to\omega)$.

\subsection{Quantum dot system}
\label{FDT-Double}

We consider now tunneling through a QDS.
In this case the problem is more complicated:
In general, the two currents $\hat I_l$ are not independent,
because $[\hat I_1,\hat I_2]\neq 0$, and thus all
correlators $S_{ll'}$ are nontrivial.
In particular, it has been proven in
Ref.~\onlinecite{LS} that the cross-correlations
${\rm Im} S_{12}(\omega)$ are sharply peaked at the  frequencies
$\omega=\Delta_{\pm}$, which is  caused by
a virtual charge-imbalance on the QDS during the cotunneling
process. The charge accumulation on the QDS for a time of order 
$\Delta_{\pm}^{-1}$ leads to an additional contribution to the noise
at finite frequency $\omega$. Thus, we expect 
that for $\omega\sim\Delta_{\pm}$ the correlators
$S_{ll'}$ cannot be expressed through the steady-state
current $I$ only and thus $I$ has to be complemented by some 
other dissipative counterparts, such as differential conductances 
$G_{ll'}$ (see Sec.~\ref{FDT-Single}).

On the other hand, at low enough frequency, $\omega \ll \Delta_{\pm}$,
the charge conservation on the QDS requires 
$\delta I_s=(\delta I_2+\delta I_1)/2\approx 0$.
Below we concentrate on the limit of low frequency
and neglect contributions of order of $\omega/\Delta_{\pm}$ to the noise power.
In  Appendix \ref{A} we prove that $S_{ss}\sim (\omega/\Delta_{\pm})^2$,
and this allows us to redefine the current and the noise power as
$I\equiv I_d=(I_2-I_1)/2$ and $S(\omega)\equiv S_{dd}(\omega)$.\cite{comment4}
In addition we require that the QDS is in the cotunneling regime,
i.e. the temperature is low enough, $k_BT\ll \Delta_{\pm}$, although
the bias $\Delta\mu$ is arbitrary (i.e.\ it can be of the order of the energy
cost) as soon as the sequential tunneling to the dot is forbidden, $\Delta_{\pm}>0$.
In this limit the current through a QDS arises due to the direct
hopping of an electron from one lead to another (through a virtual state
on the dot) with an amplitude which depends on the energy cost $\Delta_{\pm}$ 
of a virtual state. Although this process can change the state
of the QDS, the fast energy relaxation 
in the weak cotunneling regime, $w\ll w_{\rm in}$, immediately
returns it to the equilibrium state (for the opposite case,
see Secs.~\ref{microscopic}-\ref{continuum}). This allows
us to apply a perturbation expansion with respect to tunneling $V$
and to keep only first nonvanishing contributions, which we do next.

It is convenient to introduce the notation
$\bar D_l(t)\equiv\int_{-\infty}^{t}dt'\, D_l(t')$.
We notice that all relevant matrix elements,
$\langle N| D_l(t)|N+1\rangle\sim e^{-i\Delta_{+}t}$,
$\langle N-1| D_l(t)|N\rangle\sim e^{i\Delta_{-}t}$,
are fast oscillating functions of time. Thus, under the above
conditions we can write $\bar D_l(\infty)=0$,
and even more general, $\int_{-\infty}^{+\infty}dt\, D_l(t)e^{\pm i\omega t}=0$
(note that we have assumed earlier that $\omega\ll \Delta_\pm$).
Using these equalities and the cyclic property
of the trace we obtain the following result (for details of the derivation,
see Appendix~\ref{A}),
\begin{eqnarray}
&&I=e\int\limits_{-\infty}^{\infty}dt\,\langle \left[B^{\dag}(t),B(0)\right]\rangle,
\label{DB-current}\\
&&B=D_2\bar D^{\dag}_1+D^{\dag}_1\bar D_2\,.
\label{DB-amplitude}
\end{eqnarray}
Applying a similar procedure (see Appendix A), we arrive at the following
expression for the noise power $S=S_{22}$, see Eq.~(\ref{current-noise}),
\begin{equation}
S(\omega)=e^2\int\limits^{\infty}_{-\infty}dt\,
\cos (\omega t)\langle\{B^{\dag}(t),B(0)\}\rangle\,.
\label{DB-noise}
\end{equation}
where we have dropped a small contribution of order
$\omega/\Delta_{\pm}$.

Thus, we have arrived at Eqs.~(\ref{DB-current}) and
(\ref{DB-noise}) which are formally equivalent to 
Eqs.~(\ref{SB-current3}) and (\ref{SB-noise1}). Similarly to $A$
in the single-barrier case, the operator $B$ plays the role
of the effective tunneling amplitude, which annihilates
an electron in lead 1 and creates it in lead 2.
Similar to Eqs.\ (\ref{spectral}), (\ref{SB-current4}),
and (\ref{SB-noise2}) we can express the current and the noise power
\begin{eqnarray}
&&  I(\Delta\mu)
=e\tanh\left[\frac{\Delta\mu}{2k_BT}\right]{\cal B}(\Delta\mu)\,,
\label{DB-current2} \\
&& S(\omega,\Delta\mu)=\frac{e^2}{2}\sum_{\pm}
{\cal B}(\Delta\mu\pm\omega)\,,
\label{DB-noise2}
\end{eqnarray}
in terms of the spectral function
\begin{eqnarray}
&&{\cal B}(\omega) =
2\pi \sum_{ {\bf n}, {\bf m}}(\rho_{{\bf n}} +\rho_{{\bf m}})
|\langle {\bf m}|B|{\bf n}\rangle|^2\nonumber\\
&&\qquad\qquad\qquad\times\delta(\omega+E_{{\bf n}}-E_{{\bf m}})\,.
\label{spectral2}
\end{eqnarray}
The difference, however, becomes 
obvious if we notice that in contrast to the operator $A$ 
[see Eq.~(\ref{SB-tunneling})] which is a product of two fermionic 
Schr\"odinger operators 
with an equilibrium spectrum,
the operator $B$ contains an additional time integration with the time evolution
governed by $H_0=K+\sum_l\mu_lN_l$. Applying a further spectral
decomposition to the
operator $B$ [given by Eq.~(\ref{DB-amplitude})] we arrive at the expression
\begin{eqnarray}
i\langle {\bf m}|B|{\bf n}\rangle  & = &
\sum_{{\bf n'}}\frac{\langle {\bf m}| D_2|{\bf n'} \rangle
\langle {\bf n'}|D^{\dag}_1|{\bf n} \rangle }
{E_{{\bf n'}}-E_{{\bf n}}-\mu_1}  
\nonumber \\
&+&
\sum_{{\bf n''}}\frac{\langle {\bf m}|D^{\dag}_1|{\bf n''} \rangle 
\langle {\bf n''}| D_2|{\bf n} \rangle}
{E_{{\bf n''}}-E_{{\bf n}}+\mu_2}\,,
\label{spectral3}
\end{eqnarray}
where the two sums over ${\bf n'}$ and ${\bf n''}$ on the {\em lhs} are 
different by the order of tunneling sequence in the cotunneling process. 
Thus we see that the current and the  noise power depend on both
chemical potentials $\mu_{1,2}$ separately (in contrast
to the one-parameter dependence for a single-barrier junction,
see Sec.~\ref{FDT-Single}), and therefore
the shift of $\Delta\mu$ in Eq.~(\ref{DB-noise2}) by $\pm\omega$
will also shift the energy denominators of the matrix
elements on the {\em lhs} of Eq.~(\ref{spectral3}).
However, since the energy denominators are of order $\Delta_{\pm}$ the 
last effect
can be neglected and we arrive at the final result
\begin{eqnarray}
 S(\omega,\Delta\mu)=\frac{e}{2}\sum_{\pm}
\coth\left[\frac{\Delta\mu\pm\omega}{2k_BT}\right]
&&I(\Delta\mu\pm\omega)\nonumber\\
&&+ O(\omega/\Delta_{\pm})\,.
\label{DB-FDT}
\end{eqnarray}
This equation represents our nonequilibrium FDT for the 
transport through a QDS in the weak cotunneling regime.
A special case with $T, \omega=0$, giving $S=eI$, has
been derived in Ref.~\onlinecite{LS}.
To conclude this section we would like to list again the conditions
used in the derivation.
The universality of noise to current relation 
Eq.~(\ref{DB-FDT}) proven here is valid in the regime in which
it is sufficient to keep the  first nonvanishing order in the tunneling $V$
which contributes to transport and noise. 
This means that  the QDS is in the weak cotunneling regime
with $\omega, k_BT \ll \Delta_{\pm }$, and 
$w_{\rm in}\gg w$.

\section{Microscopic theory of strong cotunneling}
\label{microscopic}

\subsection{Formalism}

In this section, we give a systematic microscopic derivation of the master
equation, Eq.~(\ref{MasterEq}),
the average current, Eq.~(\ref{AvCurrent}), and the current correlators,
Eqs.~(\ref{NDnoise-all})-(\ref{correlator-delta-s}) for the QDS 
coupled to leads, as introduced
in Eqs.~(\ref{Hamiltonian})-(\ref{tunneling}), in the
strong cotunneling regime, $w_{\rm in}\ll w$.
Under this assumption the intrinsic relaxation in the QDS is very slow and will in fact
be neglected.  Thermal equilibration can only take place via coupling 
to the leads, see Sec.~\ref{me}.
Due to this slow relaxation in the QDS we find that
there are non-Poissonian correlations $\Delta S$ in the current through
the QDS because the QDS has a ``memory''; the state of the QDS after the
transmission of one electron influences the transmission of the next electron. 
A basic assumption for the following procedure is that
the system and bath are coupled only weakly and only via the perturbation
$V$, Eq.~(\ref{tunneling}).  The interaction part $H_{\rm int}$ of the 
unperturbed Hamiltonian $H_0$, Eq.~(\ref{Hamiltonian}), must therefore
be separable into a QDS and a lead part,
$H_{\rm int} = H_S^{\rm int}+H_L^{\rm int}$.
Moreover, $H_0$ conserves the number of electrons in the leads,
$[H_0,N_l]=0$, where $N_l=\sum_{k} c_{lk}^\dagger c_{lk}$.

We assume that in the distant past, $t_0\rightarrow -\infty$, 
the system is in an equilibrium state
\begin{equation}\label{initial}
  \rho _0 = \rho_{S}\otimes\rho_L,\quad
  \rho_L = \frac{1}{Z_L}e^{-K_L/k_B T},
\end{equation}
where $Z_L={\rm Tr}\, \exp[-K_L/k_B T]$, $K_L=H_L-\sum_l\mu_l N_l$,
and $\mu_l$ is the chemical potential of lead $l$.
Note that both leads are kept at the same temperature $T$.
Physically, the product form of $\rho_0$ in Eq.~(\ref{initial}) describes the
absence of correlations between the QDS and the leads in the initial state at $t_0$.
Furthermore, we assume that the initial state $\rho_0$
is diagonal in the eigenbasis of $H_0$, i.e. that the initial state is
an incoherent mixture of eigenstates of the free Hamiltonian.

In systems which can be divided into a (small) system (like the QDS) and a 
(possibly large) external ``bath'' at thermal equilibrium (here, the leads
coupled to the QDS) it turns out to be very useful to make use of the
superoperator formalism,\cite{FS,CL,LSsuper}
and of projectors $P_T=\rho_L {\rm Tr}_L$, which project
on the ``relevant'' part of the density matrix. 
We obtain $P_T\rho$ by
taking the partial trace ${\rm Tr}_L$ of $\rho$ with respect to the leads
and taking the tensor product of
the resulting reduced density matrix with the equilibrium state $\rho_L$.
Here, we will consider the projection operators
\begin{equation} 
  \label{projectors}
  P=(P_D P_N\otimes 1_L) P_T,\quad\quad Q=1-P,
\end{equation}
satisfying $P^2=P$, $Q^2=Q$, $PQ=QP=0$,
where $P$ is composed of $P_T$ and two other projectors\cite{LSsuper} $P_D$
and $P_N$, where $P_D$ projects on operators diagonal in the eigenbasis
\{$|n\rangle$\} of $H_S$, i.e. $\langle n|P_D A|m\rangle = \delta_{nm}\langle n|A|m\rangle$,
and $P_N$ projects on the subspace with $N$ particles in the QDS.
The particle number $N$ is defined by having minimal energy in
equilibrium (with no applied bias); all other particle numbers
have energies larger by at least the energy deficit\cite{cotunneling1}
$\Delta$. 
Above assumptions about the initial state Eq.~(\ref{initial}) of the
system at $t_0\rightarrow -\infty$ can now be rewritten as
\begin{equation}
  \label{initial-cond}
P\rho _0=\rho _0.
\end{equation}

For the purpose of deriving the master equation we take the Laplace transform
of the time-dependent density matrix Eq.~(\ref{dmatrix}), with the result
\begin{equation}
  \label{evolution-L}
  \rho(z) =  R(z)\rho_0.
\end{equation}
Here, $R(z)$ is the resolvent of the Liouville operator $L$, i.e.
the Laplace transform of the propagator $\exp(-itL)$,
\begin{equation}
  \label{resolvent}
  R(z) = \int_0^{\infty} \!\!\!\! dt\, e^{it(z-L)}
       = i(z-L)^{-1} \equiv \frac{i}{z-L},
\end{equation}
where $z=\omega+i\eta$. We choose $\eta>0$ in order to ensure convergence
($L$ has real eigenvalues) and at the end of the calculation take the
limit $\eta\rightarrow 0$.
We can split the resolvent into four parts by multiplying it with
the unity operator $P+Q$ from the left and the right,
\begin{equation}
  \label{res-split}
  R = PRP + QRQ + PRQ + QRP.
\end{equation}
Inserting the identity operator $-i(z-L)R(z) = -i(z-L)(P+Q)R(z)$ between
the two factors on the left hand side of $QP=0$,
$PQ=0$, $Q^2=Q$, and $P^2=P$, we obtain
\begin{eqnarray}
  QR(z)P &=&   Q\frac{1}{z-QLQ}QL_VPR(z)P,\label{resolvent-qp}\\
  PR(z)Q &=&   -iPR_0(z)PL_VQR(z)Q,\label{resolvent-pq}\\
  QR(z)Q &=&   Q\frac{i}{z-QLQ+iQL_VPR_0(z)PL_VQ}Q,\label{resolvent-qq}\\
  PR(z)P &=&   P\frac{i}{z-\Sigma(z)}P,\label{resolvent-pp}
\end{eqnarray}
where we define the {\em self-energy} superoperator
\begin{equation}
  \label{self-energy}
  \Sigma(z)=PL_VQ\frac{1}{z-QLQ}QL_VP,
\end{equation}
and the free resolvent $R_0(z)=i(z-L_0)^{-1}$.
Here, we have used the identities
\begin{eqnarray}
{\rm Tr}_L(c_{lk}\rho_L)={\rm Tr}_L(c_{lk}^\dagger\rho_L) &=& 0,
\label{property-0}\\
P_T L_V P_T = P_T \hat I_l P_T &=& 0,\label{property-1}\\
{[P,L_0]=[Q,L_0]} &=& 0,\label{property-2}\\ 
L_0P = PL_0 &=& 0.\label{property-3}
\end{eqnarray}
Equation (\ref{property-1}) follows from Eq.~(\ref{property-0}),
while Eq.~(\ref{property-2}) holds because $H_0$ neither mixes the QDS
with the leads nor does it change the diagonal elements or the
particle number of a state.
Finally, Eq.~(\ref{property-3}) can be shown with Eq.~(\ref{property-2})
and using that $P$ contains $P_D$.
For an expansion in the small perturbation $L_V$ in Eqs.~(\ref{resolvent-qp}),
(\ref{resolvent-qq}) and (\ref{self-energy}) we use the von Neumann series
\begin{eqnarray}
  \label{pert-series}
  \frac{1}{z-QLQ}Q &=& \frac{1}{z-L_0-QL_VQ}Q \nonumber\\
     &=& -iR_0(z)Q\sum_{n=0}^\infty \left[-iL_VR_0(z)Q\right]^n.
\end{eqnarray}

\subsection{Master Equation}\label{me}

Using Eqs.~(\ref{initial-cond}), (\ref{evolution-L}), and (\ref{resolvent-pp})
the diagonal part of the reduced density matrix
$\rho_S(z)=P_D P_N {\rm Tr}_L \rho(z)$ can now be written as  
\begin{equation}\label{reduced-dmatrix}
  \rho_S(z) = {\rm Tr}_L PR(z)P\rho_0
            = \frac{i}{z-\Sigma(z)}\rho_S.
\end{equation}
This equation leads to
$\dot{\rho}_S(z)=-iz\rho_S(z)-\rho_S=-i\Sigma(z)\rho_S(z)$.
The probability $\rho_n(z)=\langle n|\rho_S(z)|n\rangle$ for the QDS being in
state $|n\rangle$ then obeys the equation 
\begin{eqnarray}
  \dot\rho_n(z) &=& \sum_m W_{nm}(z) \rho_m(z),\label{master-3}\\
   W_{nm}(z) &=& -i{\rm Tr}_S \, p_n \Sigma(z) p_m  = -i\Sigma_{nn|mm}(z),
    \label{master-4}
\end{eqnarray}
with $p_n = |n\rangle\langle n|$, which is a closed equation for the density
matrix in the subspace defined by $P$ (with fixed $N$). 
In the cotunneling regime\cite{cotunneling1}, the sequential tunneling 
contribution (second order in $L_V$) to Eq.~(\ref{master-4}) vanishes.
The leading contribution [using Eqs.~(\ref{self-energy}) and
(\ref{pert-series})] is of fourth order in $L_V$,   
\begin{equation}
  \label{master-5}
  W_{nm} = {\rm Tr} \, p_n (L_V Q R_0)^3 L_V p_m \rho_L.
\end{equation}
Note that since we study the regime of small frequencies
${\rm Re}\, z = \omega \ll ||L_0Q|| \approx |E_{\bf n} - E_{\bf m}|$, where
${\bf m}\neq {\bf n}$, we can take the limit $\omega\rightarrow 0$ here.
In addition to this, we have assumed fast relaxation in the leads
and have taken the Markovian limit $z=i\eta\rightarrow 0$,
i.e.\ we have replaced $W_{nm}(z)$ in Eq.~(\ref{master-4}) by
$W_{nm}\equiv\lim_{z\rightarrow 0}W_{nm}(z)$ in Eq.~(\ref{master-5}).
The trace of $\rho$ is preserved under the time evolution Eq.~(\ref{master-3})
since $\sum_n W_{nm}$ has the form ${\rm Tr} P_N L_V A = {\rm Tr}\,[V,A] -
{\rm Tr}\,Q_N[V,A]$ where the first term vanishes exactly and the
second term invloving $Q_N=1-P_N$ is $O(\kappa)$.
After some calculation, we find that $W_{nm}$ is of the form
\begin{equation}
  \label{small-w}
  W_{nm} = w_{nm} - \delta_{nm}\sum_{m'}w_{m'n},
\end{equation}
with $w_{nm}>0$ for all $n$ and $m$.
Substituting this equation into Eq.~(\ref{master-3}) we can rewrite
the master equation in the manifestly trace-preserving
form $\dot\rho_n(z) = \sum_m [w_{nm}\rho_m(z) - w_{mn}\rho_n(z)]$,
or in real time,
\begin{equation}
  \label{MasterEq}
  \dot\rho_n(t) = \sum_m\left[ w_{nm}\rho_m(t) - w_{mn}\rho_n(t)\right].
\end{equation}
This ``classical'' master equation describes the dynamics of the QDS,
i.e.\ it describes the rates with which the probabilities $\rho_n$ for
the QDS being in state $|n\rangle$ change.
After some algebra (retaining only\cite{cotunneling2} $O(\kappa^0)$, 
cf.\ App.~\ref{B}), we find
\begin{equation}
  \label{master-6}
  w_{nm} = w^+_{nm} + w^-_{nm} + w^0_{nm},
\end{equation}
where (in the cotunneling regime)
\begin{eqnarray}
w^+_{nm} &=& w_{nm}(2,1), \quad
w^-_{nm} = w_{nm}(1,2),\label{w-pm}\\
w^0_{nm} &=& \sum_{l=1,2} w_{nm}(l,l),\label{w-zero}
\end{eqnarray}
with the ``golden rule'' rate from lead $l$ to lead $l'$,
\begin{eqnarray}
w_{nm}(l',l) &=& 2\pi \sum_{\bar{m},\bar{n}}
   |\langle{\bf n}|(D^\dagger_l, D_{l'})|{\bf m}\rangle|^2
\nonumber\\
&&\qquad\qquad\times
\delta (E_{\bf m} - E_{\bf n}-\Delta\mu_{ll'})  \rho_{L,\bar{m}}.
\label{golden-rule}
\end{eqnarray}
In this expression, $\Delta\mu_{ll'} = \mu_l-\mu_{l'}$ denotes 
the chemical potential drop between lead $l$ and lead $l'$, and
$\rho_{L,\bar{m}}=\langle \bar{m}|\rho_L|\bar{m} \rangle $.
We have defined the second order hopping operator
\begin{eqnarray}
(D^\dagger_l, D_{l'}) &=& D^\dagger_l R_0 D_{l'} +  D_{l'} R_0 D^\dagger_l\nonumber\\
       &&\qquad\qquad= -( D^\dagger_l \bar{D}_{l'} + D_{l'} \bar{D}^\dagger_l),
\label{matrix-element}
\end{eqnarray}
where $D_l$ is given in Eq.~(\ref{tunneling}),
$\bar{D_l}=\int_{-\infty}^0 D_l(t)dt$.
Note, that $(D^\dagger_l, D_{l'})$ is the amplitude of cotunneling 
from the lead $l$ to the lead $l'$ (in particular, we can write 
$B=-(D^\dagger_1, D_2)$, see Eq.~(\ref{DB-amplitude})). 
The combined index ${\bf m}=(m,\bar{m})$ contains both the QDS index
$m$ and the lead index $\bar{m}$. Correspondingly,
the basis states used above are $|{\bf m}\rangle = |m\rangle |\bar{m}\rangle$
with energy $E_{\bf m}=E_m+E_{\bar{m}}$,
where $|m\rangle$ is an eigenstate of $H_S+H_S^{\rm int}$ 
with energy $E_m$, and $|\bar{m}\rangle$ is an eigenstate of
$H_L+H_L^{\rm int}-\sum_l \mu_l N_l$ with energy $E_{\bar{m}}$.
The terms $w^\pm_{nm}$ account for the change of state in the QDS due to
a current going from lead 1 to 2 (2 to 1).
In contrast to this, the cotunneling rate $w^0_{nm}$ involves either
lead $1$ or lead $2$ and, thus, it does not contribute directly to transport.
However, $w^0_{nm}$ contributes to thermal equilibration of the QDS
via particle-hole excitations in the leads and/or QDS
(see Secs.~\ref{cold} and \ref{hot}).

\subsection{Stationary State}

In order to make use of the standard Laplace transform for finding the
stationary state $\bar\rho$ of the system, we shift the initial state to
$t_0=0$ and define the stationary state as
$\bar\rho = \lim_{t\rightarrow\infty} \rho(t)
= \lim_{t\rightarrow\infty} e^{-iLt}\rho_0$.
This can be expressed in terms of the resolvent,
\begin{equation}
  \label{dmatrix-l}
  \bar\rho = -i \lim_{z\rightarrow 0} z R(z)\rho_0,
\end{equation}
using the property
$\lim_{t\rightarrow\infty}f(t)=-i\lim_{z\rightarrow 0}zf(z)$
of the Laplace transform.  
The stationary state $\bar\rho_S$ of the QDS can be obtained
in the same way from Eq.~(\ref{reduced-dmatrix}),
\begin{equation}
  \label{reduced-stationary}
  \bar\rho_S = \lim_{z\rightarrow 0} \frac{z}{z-\Sigma(z)} \rho_S.
\end{equation}
Multiplying both sides with $z-\Sigma(z)$ and taking the limit
$z\rightarrow 0$, we obtain the condition
\begin{equation}
\label{stationary-cond}
\Sigma_0 \bar\rho_S = 0,
\end{equation}
where $\Sigma_0=\lim_{z\rightarrow 0} \Sigma(z)$.
Using Eq.~(\ref{master-4}),
this condition for the stationary state can also be expressed in
terms of $W_{nm}$,
\begin{equation}
  \label{stationary-cond-2}
  \sum_m W_{nm}\bar\rho_m = \sum_m(w_{nm}\bar\rho_m - w_{mn}\bar\rho_n)=0,
\end{equation}
which is obviously the stationarity condition for the
master equation, Eq.~(\ref{MasterEq}).

\subsection{Average Current}

The expectation value $I_l(t) = {\rm Tr} \,\hat I_l \rho(t)$
of the current $\hat I_l$ in lead $l$ [Eq.~(\ref{current-noise})]
can be obtained via its Laplace transform
\begin{equation}
  \label{expval-1}
  I_l(z)   = {\rm Tr} \, \hat I_l \rho(z) 
                          = {\rm Tr} \, \hat I_l (P+Q) R(z)P\rho_0,
\end{equation}
where we have inserted $P+Q=1$ and used 
Eqs.~(\ref{initial-cond}) and (\ref{evolution-L})
for $\rho(z)$.
According to Eq.~(\ref{property-1}) the first term vanishes.
The second term can be rewritten using
Eqs.~(\ref{resolvent-qp}) and (\ref{reduced-dmatrix}),
with the result
\begin{eqnarray}
 I_l(z)   
    &=&   {\rm Tr}\, \hat I_l Q\frac{1}{z-QLQ}QL_V \rho_S(z)\rho_L\nonumber\\
   &=&   {\rm Tr}_S\, W^I(z)\rho_S(z)
     =    \sum_{nm} W^I_{nm}(z)\rho_m(z).\label{expval-2}
\end{eqnarray}
Using the projector method, we have thus managed to express the expectation value
of the current (acting on both the QDS and the leads) in terms of
the linear superoperator $W^I$ which acts on the {\em reduced} QDS density
matrix $\rho_S$ only.
Taking $z\rightarrow 0$ in Eq.~(\ref{expval-2}), the average current in
lead $l$ in the stationary limit becomes
\begin{equation}
  \label{current-1}
  I_l  
  = \lim_{z\rightarrow 0}{\rm Tr}\, \hat I_l Q\frac{1}{z-QLQ}QL_V \bar\rho_S\rho_L.
\end{equation}
Up to now this is exact, but next
we use again the perturbation expansion Eq.~(\ref{pert-series}).
In the cotunneling regime\cite{cotunneling1,cotunneling2},
i.e.\ {\em away} from resonances, the second-order tunneling current
\begin{equation}
  \label{current-seq}
I_l^{(2)} = -i {\rm Tr} \, \hat I_l R_0 L_V\bar\rho_S\rho_L
\end{equation}
is negligible [$O(\kappa)$], and the leading contribution is the 
cotunneling current
\begin{equation}
  \label{current-2}
  I_l^{(4)} = i{\rm Tr}\, \hat I_l (Q R_0 L_V)^3 \bar\rho_S\rho_L.
\end{equation}
After further calculation we find in leading order (cf.\ App.~\ref{B})
\begin{eqnarray}
  I_2 &=& - I_1 = e\sum_{mn}w^I_{nm}\bar\rho_m,
    \label{AvCurrent}\\
  w^I_{nm} &=& w^+_{nm} - w^-_{nm},\label{w-current}
\end{eqnarray}
where $w^\pm_{nm}$ are defined in Eq.~(\ref{w-pm}).
Note again that $w^0_{nm}$ in Eq.~(\ref{w-zero})
does not contribute to the current directly,
but indirectly via the master equation Eq.~(\ref{stationary-cond-2})
which determines $\bar\rho_m$
(note that  $\bar\rho_m$ is non-perturbative in $V$).
We finally remark that for Eqs.~(\ref{current-1})-(\ref{AvCurrent}) we
do not invoke the Markovian approximation.

\subsection{Current Correlators}\label{microscopic_correlators}

Now we study the current correlators in the stationary limit.
We let $t_0\rightarrow -\infty$, therefore $\rho(t=0)\rightarrow \bar\rho$.
The symmetrized current correlator [cf.\ Eq.~(\ref{current-noise})],
\begin{equation}
  \label{correlator-t}
  S_{ll'}(t) = {\rm Re}\,{\rm Tr}\, \delta I_l(t)\delta I_{l'}\bar\rho,
\end{equation}
where $\delta I_l = \hat I_l - I_l$,
can be rewritten using the cyclic property of the trace as
\begin{equation}
  \label{correlator-t2}
  S_{ll'}(t) = {\rm Re}\, {\rm Tr}\, \delta I_l e^{-itL} \delta I_{l'} \bar\rho ,
\end{equation}
where $e^{-itL}$ acts on everything to its right.
Taking the Laplace transform and using Eq.~(\ref{dmatrix-l}) for the stationary
state $\bar\rho$, we obtain
\begin{equation}
  \label{correlator-z}
  S_{ll'}(z) = \lim_{z'\rightarrow 0}{\rm Re}(-iz'){\rm Tr}\, \delta I_l R(z)
                                        \delta I_{l'} R(z') P\rho_0, 
\end{equation}
where $z=\omega+i\eta$ and $\eta\rightarrow 0+$.
We insert $P+Q=1$ twice and use Eq.~(\ref{property-0}) with the result
\begin{equation}
  \label{correlator-z2}
  S_{ll'}(z) = S_{ll'}^{P}(z) + S_{ll'}^{Q}
                  - (i/z) I_l I_{l'},
\end{equation}
where $S_{ll'}^{Q} = S_{ll'}^{QQ} + S_{ll'}^{QP}$.
We further evaluate the contributions to $S_{ll'}(z)$ using
Eqs.~(\ref{resolvent-qp}) and (\ref{reduced-stationary}), and we obtain
\begin{equation}
  S_{ll'}^{P}(z)  = {\rm Re}\,{\rm Tr}\,\hat I_l R_Q L_V
                      P R(z) P \hat I_{l'} R_Q L_V \bar\rho, \label{corr-1}
\end{equation}
where
$R_Q = \lim_{z\rightarrow 0} (z-QLQ)^{-1}$,
and
\begin{eqnarray}
  S_{ll'}^{QQ} &=& -{\rm Re}\,{\rm Tr}\, \hat I_l R_0 L_VQR_0\hat I_{l'} R_0L_V \bar\rho \nonumber\\
                   &&-{\rm Re}\,{\rm Tr}\, \hat I_l R_0 \hat I_{l'}QR_0L_V R_0L_V \bar\rho,\label{corr-2}\\
  S_{ll'}^{QP} &=& -{\rm Re}\,{\rm Tr}\, \hat I_l R_0 L_VQR_0L_V R_0\hat I_{l'} \bar\rho.\label{corr-3}
\end{eqnarray}
While $S_{ll'}^P(z)$ as given in Eq.~(\ref{corr-1}) is a non-perturbative
result, we have used Eq.~(\ref{pert-series}) to find the leading contribution
in the tunneling amplitude $T_{lkp}$ for $S_{ll'}^{QQ}$ and $S_{ll'}^{QP}$ in
Eqs.~(\ref{corr-2}) and (\ref{corr-3}).
Also note that $QR(z)Q$ was replaced by $QR_0Q$ in Eqs.~(\ref{corr-2}) and (\ref{corr-3}),
since $\omega\ll |E_{\bf n}-E_{\bf m}|$ for ${\bf n}\neq {\bf m}$ and therefore
$S_{ll'}^{QQ}$ and $S_{ll'}^{QP}$ do not depend on $z$, i.e.\ they do not depend
on the frequency $\omega$.

In order to analyze Eq.~(\ref{corr-1}) further, we insert the resolution of unity 
$\sum_m p_m=1_S$ next to the $P$ operators in Eq.~(\ref{corr-1}) with the result
$S_{11}^P = S_{22}^P = -S_{12}^P = -S_{21}^P$ where
\begin{equation}
  \label{correlator-p-1}
  S_{11}^P = \Delta S + (i/z) I_1^2,
\end{equation}
with the non-Poissonian part
\begin{equation}
  \label{correlator-p-2}
  \Delta S(z)
    = e^2 \!\!\!\!\!\!\sum_{n,m,n',m'} w^I_{nm} \delta \rho_{mn'}(z)w^I_{n'm'}\bar\rho_{m'}.
\end{equation}
The conditional density matrix is defined as
\begin{eqnarray}
  \delta\rho_{nm}(z) &=& \rho_{nm}(z)-(i/z)\bar\rho_n,\label{delta-rho-z}\\
  \rho_{nm}(z) &=& {\rm Tr} \, p_n R(z) p_m \rho_L.\label{cond-dm}
\end{eqnarray}
Eq.~(\ref{reduced-dmatrix}) shows that $\rho_{nm}(z)$ must be a solution
of the master equation Eq.~(\ref{MasterEq}) for the
initial condition $\rho_S(0)=p_m$, or $\rho_n(0)=\delta_{nm}$.
We now turn to the remaining contribution $S_{ll'}^Q$
to $S_{ll'}(z)$ in Eq.~(\ref{correlator-z2}).
The Fourier transform $S^{\rm FT}_{ll'}(\omega)$ of the noise spectrum can be obtained
from its Laplace transform $S^{\rm LT}_{ll'}(z)$ by symmetrizing the latter,
\begin{equation}
  \label{lt-ft}
  S_{ll'}^{\rm FT}(\omega) = S_{ll'}^{\rm LT}(\omega) + S_{l'l}^{\rm LT}(-\omega).
\end{equation}
We find $S_{11}^Q = S_{22}^Q = -S_{12}^Q = -S_{21}^Q\equiv S^Q$, where
\begin{equation}
  \label{correlator-q}
   S^Q = e^2\sum_{mn}(w^+_{nm}+w^-_{nm})\bar\rho_m.
\end{equation}
Finally, we can combine Eqs.~(\ref{correlator-p-2}) and (\ref{correlator-q}),
using Eq.~(\ref{correlator-z2})
and we obtain the final result for the current correlators,
\begin{eqnarray}
  S_{11}(\omega) &=& S_{22}(\omega) = -S_{12}(\omega) = -S_{21}(\omega) \equiv S(\omega),\label{NDnoise-all}\\
  S(\omega) &=&  e^2\sum_{mn}(w^+_{nm}+w^-_{nm})\bar\rho_m + \Delta S(\omega),
             \label{NDnoise}\\
  \Delta S(\omega)
    &=& e^2 \!\!\!\!\!\!\sum_{n,m,n',m'} w^I_{nm} \delta \rho_{mn'}(\omega) w^I_{n'm'}\bar\rho_{m'},\label{correlator-delta-s}
\end{eqnarray}
where $\delta\rho_{nm}(\omega) = \rho_{nm}(\omega)-2\pi\delta(\omega)\bar\rho_n$.
Here, $\rho_{nm}(\omega)$ is the Fourier-transformed
conditional density matrix, which is obtained from the {\em symmetrized}
solution $\rho_n(t)=\rho_n(-t)$ of the
master equation Eq.~(\ref{MasterEq}) with the initial condition
$\rho_n(0)=\delta_{nm}$.
Note that $\rho_{nm}(\omega)$ is related to the
Laplace transform Eq.~(\ref{cond-dm}) via the relation
$\rho_{nm}(\omega) = \rho_{nm}^{LT}(\omega) + \rho_{nm}^{LT}(-\omega)$.

For a few-level QDS, $\delta E\sim E_C$, with inelastic
cotunneling the noise will be non-Poissonian, since the QDS is
switching between states with different currents. An explicit result
for the noise in this case can be obtained by making further
assumptions about the QDS and the coupling to the leads, and
then evaluating Eq.~(\ref{correlator-delta-s}), see the
following sections.  For the general case, we only estimate $\Delta S$.
The current is of the order $I\sim ew$, with $w$ some typical
value of the cotunneling rate $w_{nm}$, and thus 
$\delta I\sim ew$. The time between switching from one dot-state
to another due to cotunneling is approximately $\tau\sim w^{-1}$.
The correction $\Delta S$ to the Poissonian noise can be estimated
as $\Delta S\sim\delta I^2\tau\sim e^2w$, which is of the same
order as the Poissonian contribution $e I\sim e^2 w$.
Thus the correction to the Fano factor is of order unity.
In contrast to this, we find that for  elastic cotunneling
the off-diagonal rates vanish, $w_{nm}\propto \delta_{nm}$, and
therefore $\delta\rho_{nn}=0$ and $\Delta S=0$.
Moreover, at zero temperature, either $w^+_{nn}$ or $w^-_{nn}$ must
be zero (depending on the sign of the bias $\Delta\mu$). 
As a consequence, for elastic cotunneling we find Poissonian noise,
$F=S(0)/e|I|=1$.

In summary, we have derived the master equation, Eq.~(\ref{MasterEq}),
the stationary state Eq.~(\ref{reduced-stationary}) of the QDS,
the average current, Eq.~(\ref{AvCurrent}), and the current correlators,
Eqs.~(\ref{NDnoise-all})-~(\ref{correlator-delta-s})
for the QDS system coupled to leads in the cotunneling regime
under the following assumptions.
(1) Strong cotunneling regime, $w_{\rm in}\ll w$, i.e.\
negligible intrinsic relaxation in the QDS compared to the cotunneling rate; 
(2) the weak perturbation $V$ is the only coupling between the QDS
and the leads, in particular $H_{\rm int} = H_S^{\rm int}+H_L^{\rm int}$,
where $H_S^{\rm int}$ acts on the QDS and $H_L^{\rm int}$ on the leads only;
(3) no quantum correlations (neither between the QDS and the leads nor
within the QDS or the leads) in the initial state, $\rho_0=P\rho_0$;
(4) no degeneracy in the QDS, $E_n\neq E_m$ for $n\neq m$;
(5) small frequencies, $\omega\ll |E_m-E_n|$.
For the master equation Eq.~(\ref{MasterEq}) (but not for the other
results) we have additionally used the Markovian approximation, assuming
fast relaxation in the leads compared to the tunneling rate.

\section{Cotunneling through nearly degenerate states}
\label{degenerate}

Suppose the QDS has nearly degenerate states
with energies $E_n$,  and
level spacing $\Delta_{nm}=E_n - E_m$,
which is much smaller
than the average level spacing $\delta E$. In the regime,
$\Delta\mu, k_BT, \Delta_{nm}\ll \delta E$, the only allowed
cotunneling processes
are the transitions between nearly degenerate states.
The noise power is given by Eqs.\ (\ref{NDnoise}) and
(\ref{correlator-delta-s}),
and below we calculate the correlation correction to the noise, $\Delta S$.
To proceed with our calculation
we rewrite Eq.~(\ref{MasterEq}) for $\delta\rho(t)$
(see Eq.\ (\ref{delta-rho-z}))
as a second-order differential equation in matrix form
\begin{equation}
\delta\ddot\rho(t) = W^2\delta\rho(t),\quad \delta\rho(0)=1-\bar\rho ,
\label{second-eq}
\end{equation}
where $W$ is defined in Eq.~(\ref{small-w}).
We solve this equation by
Fourier transformation,
\begin{equation}
\delta\rho(\omega)=-\frac{2W}{W^2+\omega^2 1},
\label{delta-rho}
\end{equation}
where we have used $W\bar\rho=0$.
We substitute $\delta\rho$ from this equation into
Eq.~(\ref{correlator-delta-s})
and write the result in a compact matrix form,
\begin{equation}
\Delta S(\omega)=
-e^2\sum_{n,m}\left[w^{I}\frac{2W}{W^2+\omega^2
1}w^{I}\bar\rho\right]_{nm}.
\label{result01}
\end{equation}
This equation gives the formal solution of the noise problem for
nearly degenerate states. As an example we consider a two-level system.

Using the detailed balance equation, $w_{21}\rho_1=w_{12}\rho_2$, we obtain
for
the stationary probabilities $\rho_1=w_{12}/(w_{12}+w_{21})$, and
$\rho_2=w_{21}/(w_{12}+w_{21})$. From Eq.~(\ref{AvCurrent}) we get
\begin{equation}
I=e\frac{w_{12}(w^{I}_{11}+w^{I}_{21})+w_{21}(w^{I}_{22}+w^{I}_{12})}
{w_{12}+w_{21}}.
\label{two-current}
\end{equation}
A straightforward calculation with the help of Eq.~(\ref{delta-rho})
gives for the correction to the Poissonian noise
\begin{eqnarray}
\Delta S(\omega) & = &
\frac{2e^2(w^{I}_{11}+w^{I}_{21}-w^{I}_{22}-w^{I}_{12})}
{(w_{12}+w_{21})[\omega^2+(w_{12}+w_{21})^2]}\times\nonumber \\
& \times  & \left[w^{I}_{11}w_{12}w_{21}+w^{I}_{12}w^2_{21}-
(1\leftrightarrow 2)\right].
\label{two-noise}
\end{eqnarray}
In particular, the zero frequency noise
$\Delta S(0)$
diverges if the ``off-diagonal'' rates $w_{nm}$ vanish.
This divergence has to be cut at
$\omega$, or at the relaxation rate $w_{\rm in}$ due to coupling to
the bath (since $w_{12}$ in this case
has to be replaced with $w_{12}+w_{\rm in}$).
The physical origin of the divergence is rather
transparent: If the off-diagonal rates $w_{12},w_{21}$
are small, the QDS goes into an
unstable state where it switches between states 1 and 2 with
different currents in general.\cite{comment5}
The longer the QDS stays in the
state 1 or 2 the larger the zero-frequency noise power is. However,
if $w^{I}_{11}+w^{I}_{21}=w^{I}_{22}+w^{I}_{12}$,
then $\Delta S(\omega)$ is suppressed to 0. For
instance, for the QDS in the spin-degenerate state with an odd
number of electrons $\Delta S(\omega)=0$, since the two states
$|\uparrow\rangle$ and $|\downarrow\rangle$ are physically
equivalent. The other example of such a suppression
of the correlation correction $\Delta S$
to noise is given by a multi-level QDS, $\delta E\ll E_C$,
where the off-diagonal rates are
small compared to the diagonal (elastic) rates.\cite{averinazarov}
Indeed, since the main contribution to the elastic rates comes
from transitions through many virtual states,
which do not participate in inelastic cotunneling,
they do not depend on the initial
conditions, $w^I_{11}=w^I_{22}$, and cancel in the numerator of
Eq.~(\ref{two-noise}), while they are still present in the current.
Thus the correction $\Delta S/I$ vanishes in this case.
Further below in this section we consider a few-level QDS,
$\delta E\sim E_C$, where $\Delta S\neq 0$.

To simplify further analysis we consider for a moment the case, where
the singularity in the noise is most pronounced, namely,
$\omega=0$ and $|\Delta_{12}|\ll\Delta\mu, k_BT$, so that
$w^I_{12}=w^I_{21}$, and $w_{12}=w_{21}$.
Then, from Eqs.\ (\ref{two-current}) and (\ref{two-noise})
we obtain
\begin{eqnarray}
&& I=\frac{1}{2}(I_{1}+I_{2})\,,\quad I_{n}=e\sum_{m=1,2}w^I_{mn}\,,
\label{two-current2} \\
&& \Delta S(0)=\frac{(I_{1}-I_{2})^2}{4w_{12}}\,,
\label{two-noise2}
\end{eqnarray}
where $I_n$ is the current through the $n$-th level of the QDS.
Thus in case $|\Delta_{12}|\ll\Delta\mu, k_BT$ the
following regimes have to be distinguished:
(1) If $k_BT\lesssim\Delta\mu$, then $I_n\propto\Delta\mu$,
$w_{12}\propto\Delta\mu$, and thus both, the total current
$I=e^{-1}G_D\Delta\mu$, and
the total noise $S=FG_D\Delta\mu$ are linear in the bias $\Delta\mu$
(here $G_D$ is the conductance of the QDS).
The total shot noise in this regime is super-Poissonian with the Fano factor
$F\sim I/(ew_{12})\gg 1$.
(2) In the regime $\Delta\mu\lesssim k_BT\lesssim F^{1/2}\Delta\mu$
the noise correction (\ref{two-noise2}) arises because of the thermal
switching
the QDS between two states $n=1,2$, where the currents are linear in the
bias,
$I_n\sim G_D\Delta\mu/e$. The rate of switching is $w_{12}\propto k_BT$, and
thus
$\Delta S\sim FG_D\Delta\mu^2/(k_BT)$. Since $k_BT/\Delta\mu\lesssim
F^{1/2}$, the
noise correction $\Delta S$
is the dominant contribution to the noise, and thus the total noise $S$
can be interpreted as being a thermal telegraph noise.\cite{Kogan}
(3) Finally, in the regime $F^{1/2}\Delta\mu\lesssim k_BT$
the first term on the rhs of
Eq.\ (\ref{NDnoise}) is the dominant contribution, and the total noise
becomes
an equilibrium Nyquist noise, $S=2G_Dk_BT$.

\begin{figure}
  \begin{center}
    \leavevmode
\epsfxsize=5.5cm
\epsffile{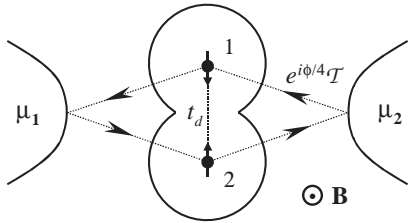}
  \end{center}
\caption{
Double-dot (DD) system containing two electrons and
being weakly coupled to metallic leads 1, 2,
each of which is at the chemical potential
$\mu_1$, $\mu_2$. The tunneling amplitudes between dots and leads are
denoted
by $\cal T$. The tunneling ($t_d$) between the dots
results in a singlet-triplet splitting $J\sim t_d^2/U$
with the singlet being a ground state.\protect\cite{Burkard}
The tunneling path between dots
and leads 1 and 2 forms a closed loop (shown by arrows)
so that the Aharonov-Bohm phase $\phi$ will be accumulated by
an electron traversing the DD.
}
\label{double-d}
\end{figure}

We notice that
for the noise power to be divergent the off-diagonal rates
$w_{12}$ and $w_{21}$ have to vanish simultaneously.
However, the matrix $w_{nm}$ is not symmetric since the off-diagonal
rates depend on the bias in a different way. On the other hand,
both rates contain the same matrix element of the cotunneling amplitude
$(D^{\dag}_l,D_{l'})$, see Eqs.~(\ref{golden-rule}) and
(\ref{matrix-element}).
Although in general this matrix element is not small, it can vanish
because of different symmetries of the two states. To illustrate this effect
we consider the transport through a double-dot (DD) system
(see Ref.\ \onlinecite{LS} for details) as an example. Two leads
are equally coupled to two dots  in such a way
that a closed loop is formed, and the dots are also connected,
see Fig.~\ref{double-d}.
Thus, in a magnetic field the tunneling is described by the Hamiltonian
Eq.~(\ref{tunneling}) with
\begin{eqnarray}
&&D_l=\sum_{s,j}T_{lj}c^{\dag}_{ls}d_{js}\,,
\qquad l,j=1,2\,,
\label{DD-D}\\
&&T_{11}=T_{22}=T^{*}_{12}=T^{*}_{21}=
e^{i\phi/4}{\cal T}\,,
\label{equal}
\end{eqnarray}
where the last equation expresses the equal coupling of dots and leads
and $\phi$ is the Aharonov-Bohm phase.
Each dot contains one electron, and weak tunneling $t_d$ between the dots
causes
the exchange splitting\cite{Burkard}
$J\sim t_d^2/U$ (with $U$ being the on-site repulsion)
between one spin singlet and three triplets
\begin{eqnarray}
&&|S\rangle=\frac{1}{\sqrt{2}}
[d^{\dag}_{1\uparrow}d^{\dag}_{2\downarrow}\!-
d^{\dag}_{1\downarrow}d^{\dag}_{2\uparrow}]|0\rangle\,,
\nonumber\\
&&|T_0\rangle=\frac{1}{\sqrt{2}}
[d^{\dag}_{1\uparrow}d^{\dag}_{2\downarrow}\!+
d^{\dag}_{1\downarrow}d^{\dag}_{2\uparrow}]|0\rangle\,,
\label{basis}\\
&&|T_+\rangle=d^{\dag}_{1\uparrow}d^{\dag}_{2\uparrow}|0\rangle\,,
\quad
|T_-\rangle=d^{\dag}_{1\downarrow}d^{\dag}_{2\downarrow}|0\rangle\,.
\nonumber
\end{eqnarray}
In the case of zero magnetic field, $\phi=0$, the tunneling Hamiltonian $V$
is symmetric with respect to the exchange of electrons, $1\leftrightarrow
2$.
Thus the matrix element of the cotunneling transition between the singlet
and three triplets $\langle S|V(E-H_0)^{-1}V|T_i\rangle$, $i=0,\pm$,
vanishes
because these states have different orbital symmetries.
A weak magnetic field breaks the symmetry, contributes to the off-diagonal
rates,
and thereby reduces noise.

\begin{figure}
  \begin{center}
    \leavevmode
\epsfxsize=8.5cm
\epsffile{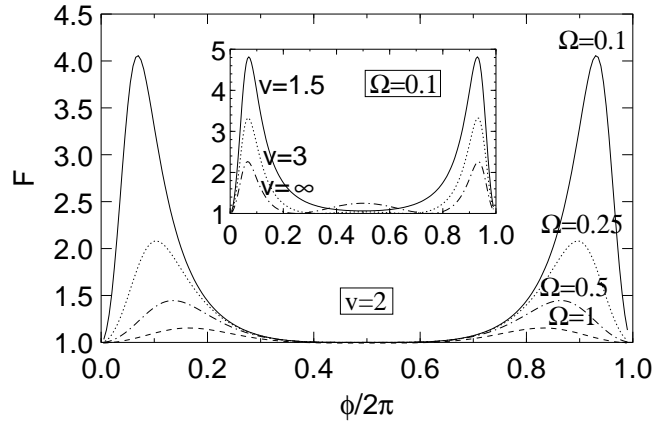}
  \end{center}
\caption{
The Fano factor $F=S(\omega)/I$, with the noise power
$S(\omega)$ given in Eqs.~(\ref{NDnoise}) and
(\ref{DD-noise}), and with the current through the DD, $I$,
given in Eqs.~(\ref{two-current}) and (\ref{DD-rates}), is
plotted as a function of the Aharonov-Bohm phase $\phi$
for the normalized bias $v\equiv \Delta\mu/J=2$ and for four
different normalized frequencies
$\Omega\equiv\omega /[G(2\Delta\mu - J)]=0.1$, $0.25$,
$0.5$, and $1$. Inset: the same, but with fixed frequency
$\Omega=0.1$, where the bias $v$ takes the values
$1.5$, $3$, and $\infty$.
}
\label{noise-corr}
\end{figure}

The fact that in the perturbation $V$ all spin indices are traced out
helps us to map the four-level system to only two states
$|S\rangle$ and $|T\rangle$ classified according to the orbital
symmetry (since all triplets are antisymmetric in orbital space).
In Appendix \ref{C} we derive the mapping to a two-level system
and  calculate the transition rates $w^{+}_{nm}$ and $w^{0}_{nm}$
($n,m=1$ for a singlet
and $n,m=2$ for all triplets)
using Eqs.~(\ref{golden-rule}) and (\ref{matrix-element})
with the operators $D_l$ given by Eq.~(\ref{DD-D}).
Doing this we obtain the following result
\begin{eqnarray}
&& w^{0}_{nm}=0,\nonumber \\
&& w^{+}_{nm}=
\frac{\pi}{2}\left(\frac{\nu{\cal T}^2}{\Delta_{-}}\right)^2 \nonumber \\
&&
\times\left\{\begin{array}{ll}
(1\!+\!\cos\phi)\Delta\mu & (1\!-\!\cos\phi)(\Delta\mu\!+\!J) \\
3(1\!-\!\cos\phi)(\Delta\mu\!-\!J) & 3(1\!+\!\cos\phi)\Delta\mu \\
\end{array}
\right\},
\label{DD-rates}
\end{eqnarray}
which holds close to the sequential tunneling peak,
$\Delta_{-}\ll\Delta_{+}\sim U$
(but still $\Delta_{-}\gg J,\Delta\mu$),
and for $\Delta\mu>J$. We substitute this equation into the
Eq.~(\ref{two-noise})
and write the correction $\Delta S(\omega)$ to the Poissonian noise
as a function of normalized bias $v=\Delta\mu/J$ and normalized
frequency $\Omega=e\omega /[G(2\Delta\mu - J)]$
\begin{equation}
\Delta S(\omega)= 6eGJ
\frac{(v^2-1)[1+(v-1)\cos\phi]^2(1-\cos\phi)}{(2v-1)^3[\Omega^2+(1-\cos\phi)
^2]},
\label{DD-noise}
\end{equation}
where $G=\pi e (\nu{\cal T}^2/\Delta_{-})^2$ is the conductance
of a single dot in the cotunneling regime.\cite{Recher}
{}From Eq.~(\ref{DD-noise}) it follows that
the noise power has singularities as a function of $\omega$ for zero
magnetic field,
and it has singularities at $\phi=2\pi m$ (where $m$ is integer) as a
function
of the magnetic field (see Fig.~\ref{noise-corr}).
We would like to emphasize that the noise is singular even if the exchange
between
the dots is weak, $J\ll \Delta\mu$. Note however, that our classical
approach, which
neglects the off-diagonal elements of the density matrix $\rho(t)$, can only
be
applied for weak enough tunneling, $w_{nm}\ll J$.
In the case $\Delta\mu<J$ the transition from the singlet
to the triplet is forbidden by conservation of energy, $w^{+}_{21}=0$,
and we immediately obtain from Eq.~(\ref{two-noise}) that $\Delta
S(\omega)=0$,
i.e.\ the total noise is Poissonian (as it is always the case for elastic
cotunneling).
In the case of large bias, $\Delta\mu\gg J$,
two dots contribute independently to the current $I=2e^{-1}G\Delta\mu$,
and from Eq.\ (\ref{DD-noise}) we obtain the Fano factor
\begin{equation}
F=\frac{3}{8}\,\frac{\cos^2\!\phi(1-\cos\phi)}{\Omega^2+(1-\cos\phi)^2},
\quad \Delta\mu\gg J.
\label{Fano}
\end{equation}
This Fano factor controls the transition to the telegraph noise and then to
the equilibrium noise at high temperature, as described above.
We notice that if the coupling of the dots to the leads is not equal, then
$w^{0}_{nm}\neq 0$ serves as a cut-off of the singularity in
$\Delta S(\omega)$.

Finally, we remark that the Fano factor is a periodic
function of the phase $\phi$ (see Fig.~\ref{noise-corr});
this is nothing but an Aharonov-Bohm effect in the
noise of the cotunneling transport through the DD.
However, in contrast to the  Aharonov-Bohm effect
in the cotunneling current through the DD which has
been discussed earlier in Ref.~\onlinecite{LS}, the noise
effect does not allow us to probe the ground state of the DD,
since the DD is already in a mixture of the singlet and three
triplet states.

\section{Cotunneling through continuum of single-electron states}
\label{continuum}

We consider now the transport through a multi-level QDS with 
$\delta E\ll E_C$. In the low bias regime, 
$\Delta\mu\ll (\delta E\, E_C)^{1/2}$,
the elastic cotunneling dominates transport,\cite{averinazarov}
and according to the results of Secs.\ \ref{microscopic} 
and \ref{degenerate}  the noise
is Poissonian. Here we consider the opposite regime of
inelastic cotunneling,
$\Delta\mu\gg (\delta E\, E_C)^{1/2}$.
Since a large number $M$ of levels participate in transport,
we can neglect the correlations which we have studied in the previous
section, since they become a $1/M$-effect. 
Instead, we concentrate on the heating effect, which
is not relevant for the 2-level system considered before. 
The condition for strong cotunneling has to be rewritten
in a single-particle form, $\tau_{\rm in}\gg\tau_c$, where
$\tau_{\rm in}$ is the single-particle energy relaxation 
time on the QDS due to the coupling
to the environment, and $\tau_c$ is the time of the cotunneling
transition, which can be estimated as $\tau_c\sim e\nu_D\Delta\mu/I$
(where $\nu_D$ is the density of QDS states).
Since the
energy relaxation rate on the QDS is small, 
the multiple cotunneling transitions can cause
high energy excitations on the dot, 
and this leads to a nonvanishing backward
tunneling, $w^{-}_{nm}\neq 0$. 
In the absence of correlations between cotunneling events, 
Eqs.~(\ref{AvCurrent}), (\ref{w-current}) and (\ref{NDnoise}) can be rewritten in terms of
forward and backward tunneling currents $I_{+}$ and $I_{-}$,
\begin{eqnarray}
&& I=I_{+}-I_{-}\,,\quad S=e(I_{+}+I_{-}),
\label{IS} \\
&&I_{\pm}=e\sum_{n,m}w^{\pm}_{nm}\bar\rho_m\,, \label{Ipm1}
\end{eqnarray}
where the transition rates are given by Eqs.~(\ref{w-pm}) and (\ref{golden-rule}).

It is convenient to rewrite the currents $I_{\pm}$ in a single-particle basis. 
To do so we substitute the rates Eq.~(\ref{golden-rule}) into
Eq.~(\ref{Ipm1}) and neglect the dependence of the
tunneling amplitudes Eq.~(\ref{tunneling}) on the quantum numbers
$k$ and $p$, $T_{lkp}\equiv T_{l}$, which is a reasonable assumption for QDS with
a large number of electrons. Then we define the distribution function on
the QDS as 
\begin{equation}
f(\varepsilon)=\nu_D^{-1} \sum_{p}\delta(\varepsilon -
\varepsilon_p){\rm Tr}\,\bar\rho d^{\dag}_{p}d_{p}
\end{equation}
and replace the summation
over $p$ with an integration over $\varepsilon$. 
Doing this we obtain the following expressions for $T=0$
\begin{eqnarray}
&& I_{\pm}  =  C_{\pm}\frac{G_1G_2}{2\pi e^3}
\left(\frac{1}{\Delta_{+}}+\frac{1}{\Delta_{-}}\right)^2(\Delta\mu)^3,
\label{Ipm2} \\
&& C_{\pm} = \frac{1}{\Delta\mu^3}
\int\!\!\int\! d\varepsilon d\varepsilon' 
\Theta(\varepsilon\! -\varepsilon'\!\pm\Delta\mu)
f(\varepsilon)[1-f(\varepsilon')],
\label{Cpm}
\end{eqnarray}
where $G_{1,2}=\pi e^2\nu\nu_D|T_{1,2}|^2$ 
are the tunneling conductances of
the two barriers,
and where we have introduced the function 
$\Theta(\varepsilon)=\varepsilon\theta(\varepsilon)$
with $\theta(\varepsilon)$ being the step-function.
In particular, using the property
$\Theta(\varepsilon+\Delta\mu)-\Theta(\varepsilon-\Delta\mu)
=\varepsilon +\Delta\mu$
and fixing
\begin{equation}
\int d\varepsilon
[f(\varepsilon)-\theta({-\varepsilon})]=0,
\label{symmetry}
\end{equation}
(since $I_{\pm}$ given by
Eq.~(\ref{Ipm2}) and Eq.~(\ref{Cpm}) do not depend on the shift
$\varepsilon\to\varepsilon+const$) we arrive at the following
general expression for the cotunneling current
\begin{eqnarray}
&&I= \Lambda\,\frac{G_1G_2}{12\pi e^3}
\left(\frac{1}{\Delta_{+}}+\frac{1}{\Delta_{-}}\right)^2(\Delta\mu)^3,
\label{I-continuum}\\
&&\Lambda=1+12\Upsilon/(\Delta\mu)^2, \label{prefactor}\\
&&\Upsilon=\int d\varepsilon \varepsilon [f(\varepsilon)-
\theta(-\varepsilon)]\geq 0, \label{Upsilon}
\end{eqnarray}
where the value $\nu_D\Upsilon$ has the physical meaning of
the energy acquired by the QDS due to the cotunneling
current through it.

We have deliberately  introduced the functions $C_{\pm}$ in the
Eq.~(\ref{Ipm2}) to emphasize the fact that if the distribution
$f(\varepsilon)$ scales with the bias $\Delta\mu$ (i.e.\ $f$ is a
function of $\varepsilon/\Delta\mu$), then $C_{\pm}$ become
dimensionless universal numbers. Thus both, the prefactor $\Lambda$
[given by Eq.~(\ref{prefactor})] in the cotunneling current, and the
Fano factor $F=S/(eI)$, where $S=eI+\Delta S_h$,
\begin{equation}
F=\frac{C_{+}+C_{-}}{C_{+}-C_{-}},
\label{F-continuum}
\end{equation}
take their universal values, which do not depend on the bias
$\Delta\mu$. We consider now such universal regimes.
The first example is the case of weak cotunneling, 
$\tau_{\rm in}\ll\tau_c$, when the QDS is in its ground state,
$f(\varepsilon)=\theta (-\varepsilon)$, and the thermal energy of
the QDS vanishes, $\Upsilon=0$. Then $\Lambda=1$, and 
Eq.~(\ref{I-continuum}) reproduces the results of Ref.\
\onlinecite{averinazarov}. As we have already mentioned, the
backward current vanishes, $I_{-}=0$, and the Fano factor acquires
its full Poissonian value $F=1$, in agreement with our 
nonequilibrium FDT proven in Sec.\ \ref{FDT-Double}. 
In the limit of strong cotunneling,
$\tau_{\rm in}\gg\tau_c$, the energy relaxation 
on the QDS can be neglected.
Depending on the electron-electron scattering time $\tau_{ee}$ two
cases have to be distinguished: The regime of cold electrons
$\tau_{ee}\gg\tau_c$ and regime of hot electrons
$\tau_{ee}\ll\tau_c$ on the QDS. Below we discuss both
regimes in detail and demonstrate their universality.

\subsection{Cold electrons}
\label{cold}

In this regime  the electron-electron scattering on the QDS can be
neglected and the distribution $f(\varepsilon)$ has to be found
from the master equation Eq.~(\ref{MasterEq}). We multiply this
equation by $\nu_D^{-1}\sum_{p}\delta(\varepsilon -
\varepsilon_p) \langle n|d^{\dag}_{p}d_{p}|n\rangle$, sum over
$n$ and use the tunneling rates from Eq.~(\ref{golden-rule}). Doing this we
obtain the standard stationary kinetic equation which can be
written in the following form
\begin{eqnarray}
&& \int d\varepsilon'\sigma(\varepsilon'
-\varepsilon)f(\varepsilon')[1-f(\varepsilon)]   \nonumber \\
&& \qquad\qquad\qquad =\int d\varepsilon' \sigma(\varepsilon
-\varepsilon')f(\varepsilon)[1-f(\varepsilon')],
\label{kineq1}\\
&& \sigma(\varepsilon)=2\lambda\Theta(\varepsilon)+
\sum_{\pm}\Theta(\varepsilon \pm\Delta\mu),
\label{kernel}
\end{eqnarray}
where $\lambda=(G_1^2+G_2^2)/(2G_1G_2)\geq1$ arises from the equilibration
rate $w^0_{mn}$, see Eq.~(\ref{w-zero}).
(We assume that if the limits of the integration over energy $\varepsilon$ are not
specified, then the integral goes from $-\infty$ to $+\infty$.) From the form of this
equation we immediately conclude that its solution is a function
of $\varepsilon/\Delta\mu$, and thus the cold electron regime is universal
as defined in the previous section.
It is easy to check that the detailed balance does not hold, and in addition
$\sigma(\varepsilon)\neq\sigma(-\varepsilon)$. Thus we face a difficult problem
of solving Eq.~(\ref{kineq1}) in its full nonlinear form.
Fortunately, there is a way to avoid this problem and to reduce the equation
to a linear form which we show next.

We group all nonlinear terms on the rhs of Eq.~(\ref{kineq1}):
$\int
d\varepsilon'\sigma(\varepsilon'-\varepsilon)f(\varepsilon')=
h(\varepsilon)f(\varepsilon)$, where $h(\varepsilon)=\int
d\varepsilon'
\left\{\sigma(\varepsilon'-\varepsilon)f(\varepsilon')+
\sigma(\varepsilon-\varepsilon')[1-f(\varepsilon')]\right\}$. The
trick is to rewrite the function $h(\varepsilon)$ in terms of
known functions. For doing this we split the integral in
$h(\varepsilon)$ into two integrals over $\varepsilon'>0$ and
$\varepsilon'<0$, and then use Eq.~(\ref{symmetry}) and the
property of the kernel $\sigma(\varepsilon)-\sigma(-\varepsilon)=
2(1+\lambda)\varepsilon$ to regroup terms in such a way that
$h(\varepsilon)$ does not contain $f(\varepsilon)$ explicitly.
Taking into account Eq.~(\ref{Upsilon}) we
arrive at the following linear integral equation
\begin{eqnarray}
&& \int d\varepsilon'\sigma(\varepsilon'-\varepsilon)f(\varepsilon')
\nonumber \\
&&\qquad\qquad\qquad
=[(1+\lambda)(\varepsilon^2+2\Upsilon)+(\Delta\mu)^2]f(\varepsilon),
\label{kineq2}
\end{eqnarray}
where the parameter
$\Upsilon$  is the only signature of the nonlinearity
of Eq.~(\ref{kineq1}).

Since Eq.~(\ref{kineq2}) represents an eigenvalue
problem for a linear operator, it can in general have more than
one solution. Here we demonstrate that there is only one physical
solution, which satisfies the conditions
\begin{equation}
0\leq f(\varepsilon)\leq 1, \quad f(-\infty)=1, \quad f(+\infty)=0.
\label{conditions}
\end{equation}
Indeed, using a standard procedure one can show that two solutions
of the integral equation (\ref{kineq2}), $f_1$ and $f_2$,
corresponding to different parameters $\Upsilon_1\neq\Upsilon_2$
should be orthogonal, $\int d\varepsilon
f_1(\varepsilon)f_2(-\varepsilon)=0$. This contradicts the
conditions Eq.~(\ref{conditions}). The solution is also unique for the
same $\Upsilon$, i.e. it is not degenerate (for a proof, see
Appendix \ref{D}). From Eq.~(\ref{kineq1}) and conditions
Eq.~(\ref{conditions}) it follows that if $f(\varepsilon)$ is a
solution then $1-f(-\varepsilon)$ also satisfies 
Eqs.~(\ref{kineq1}) and (\ref{conditions}). Since the solution is
unique, it has to have the symmetry
$f(\varepsilon)=1-f(-\varepsilon)$.

\begin{figure}
  \begin{center}
    \leavevmode
\epsfxsize=8.5cm
\epsffile{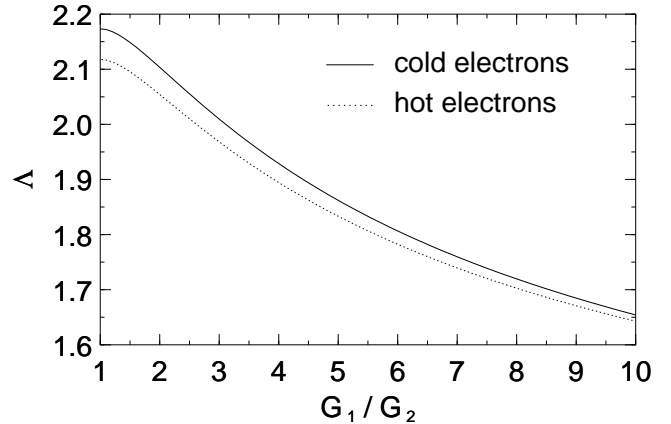}
  \end{center}
\caption{
The prefactor $\Lambda$ in the expression (\ref{I-continuum})
for the cotunneling current characterizes a universal cotunneling 
transport in the regime of weak cotunneling,
$\tau_{\rm in}\ll\tau_c$, ($\Lambda=1$, see 
Ref.\ \protect\onlinecite{averinazarov}),
and in the regime of strong cotunneling, $\tau_{\rm in}\gg\tau_c$
($\Lambda>1$). Here $\Lambda$ is
plotted as a function of $G_1/G_2$ (same as a function
of $G_2/G_1$) for the strong cotunneling,
for the cold-electron case, $\tau_{ee}\gg\tau_c$ (solid line)
and for the hot-electron case, $\tau_{ee}\ll\tau_c$ (dotted line).
$G_{1,2}$ are the tunneling conductances 
of a junctions connecting leads 1 and 2 with the QDS. 
}
\label{cold-hot}
\end{figure}

We solve Eqs.~(\ref{kineq2}) and (\ref{conditions}) numerically 
and use Eqs.\ (\ref{Cpm}) and (\ref{F-continuum}) to find that 
the Fano factor is very close to 1 
(it does not exceed the value $F\approx 1.006$).
Next we use Eqs.\ (\ref{prefactor}) and (\ref{Upsilon})
to calculate the prefactor $\Lambda$ and plot the result as 
a function of the ratio of tunneling conductances, $G_1/G_2$,
(Fig.~\ref{cold-hot}, solid line). For equal coupling
to the leads, $G_1=G_2$, the prefactor $\Lambda$ takes its maximum
value $2.173$, and thus the cotunneling current is approximately
twice as large compared to its value for the case of
weak cotunneling, $\tau_{\rm in}\ll\tau_c$. 
$\Lambda$ slowly decreases with increasing  
asymmetry of coupling and tends to its minimum value $\Lambda =1$ 
for the strongly asymmetric coupling
case $G_1/G_2 or G_2/G_1\gg 1$.

\subsection{Hot electrons}
\label{hot}

In the regime of hot electrons, $\tau_{ee}\ll\tau_c$, the
distribution is given by the equilibrium Fermi function
$f_F(\varepsilon)=\left[1+\exp(\varepsilon/k_BT_e)\right]^{-1}$, 
while the electron temperature $T_e$ has to be found
self-consistently from the kinetic equation.  
Eq.~(\ref{kineq1}) has to be modified to take into account 
electron-electron interactions. This can be done by adding the
electron collision integral $I_{ee}(\varepsilon)$ to the rhs.\
of (\ref{kineq1}).
Since the form of the distribution is known we need only the
energy balance equation, which can be derived by multiplying the
modified equation (\ref{kineq1}) by $\varepsilon$ and integrating
it over $\varepsilon$. The contribution from the collision
integral $I_{ee}(\varepsilon)$ vanishes, because the
electron-electron scattering conserves the energy of the system.
Using the symmetry $f_F(\varepsilon)=1-f_F(-\varepsilon)$ we
arrive at the following equation
\begin{equation}
\int\!\!\int d\varepsilon d\varepsilon' f_F(\varepsilon')[1-f_F(\varepsilon)]
\sigma(\varepsilon'-\varepsilon)\varepsilon =0.
\label{kineq3}
\end{equation}
Next we regroup the terms in this equation such that it
contains only integrals of the form $\int_{0}^{\infty}d\varepsilon
f_F(\varepsilon)(\ldots)$. This allows us to get rid of nonlinear
terms, and we arrive at the following equation,
\begin{equation}
\int d\varepsilon \varepsilon^3
[f_F(\varepsilon)-\theta(-\varepsilon)]
+3\Upsilon^2=\frac{(\Delta\mu)^4}{8(1+\lambda)}\,,
\label{kineq4}
\end{equation}
which holds also for the regime of cold electrons. Finally, we
calculate the integral in Eq.~(\ref{kineq4}) and express the result in
terms of the dimensionless parameter $\alpha=\Delta\mu/k_BT_e$,
\begin{equation}
\alpha=\pi\left[8(1+\lambda)/5\right]^{1/4}.
\label{alpha}
\end{equation}
Thus, since the distribution again depends on the ratio
$\varepsilon/\Delta\mu$, the hot electron regime is also
universal.

The next step is to substitute the Fermi distribution 
function with the temperature
given by Eq.~(\ref{alpha}) into Eq.~(\ref{Cpm}). 
We calculate the integrals and
arrive at the closed analytical expressions for the values of interest,
\begin{eqnarray}
&&
\Lambda=1+\frac{2\pi^2}{\alpha^2}=1+\sqrt{\frac{5}{2(1+\lambda)}}\,,
\label{I-hot}
\\ && F=1+\frac{12}{2\pi^2+\alpha^2}
\sum_{n=1}^{\infty}\left[\frac{1}{n^2} +\frac{2}{\alpha
n^3}\right]e^{-\alpha n}, \label{F-hot}
\end{eqnarray}
where again $\lambda=(G_1^2+G_2^2)/2G_1G_2\geq 1$.
It turns out that similar to the case of cold electrons,
Sec.\ \ref{cold}, the Fano factor for hot electrons is very 
close to $1$ (namely, it does not exceed the value $F\approx 1.007$). 
Therefore, we do not expect that the
super-Poissonian noise considered in this section
(i.e.\ the one which is due to heating 
of a large QDS caused by inelastic cotunneling through it) 
will be easy to observe in experiments. 
On the other hand, the transport-induced heating of a large 
QDS can be observed in the cotunneling current 
through the prefactor $\Lambda$,
which according to Eq.\ (\ref{I-hot}) takes its maximum value
$\Lambda= 1+\sqrt{5/4}\approx 2.118$ for $G_1=G_2$ and 
slowly reaches its minimum value $1$ with increasing (or decreasing) 
the ratio $G_1/G_2$ (see Fig.~\ref{cold-hot}, dotted line).
Surprisingly, the two curves of $\Lambda$ vs $G_1/G_2$ for the
cold- and hot-electron regimes lie very close, 
which means that the effect of the electron-electron scattering
on the cotunneling transport is rather weak.

\section{Conclusions}

The physics of the noise of cotunneling is discussed in the
Introduction. Here we give a short summary of our results.

In Sec.~\ref{FDT}, we have derived the non-equilibrium FDT, i.e.\ the
universal relations Eqs.~(\ref{SB-FDT}) and (\ref{DB-FDT}) between the
current and the noise, for single-barrier junctions and for QDS in the weak cotunneling regime, 
respectively. Taking the limit $T,\omega\rightarrow 0$, we
show that the noise is Poissonian, i.e.\ $F=1$. 

In Sec.~\ref{microscopic}, we have derived the master equation, Eq.~(\ref{MasterEq}),
the stationary state Eq.~(\ref{reduced-stationary}) of the QDS,
the average current, Eq.~(\ref{AvCurrent}), and the current correlators,
Eqs.~(\ref{NDnoise-all})-(\ref{correlator-delta-s})
for a non-degenerate QDS system ($E_n\neq E_m$, $n\neq m$) coupled to leads
in the strong cotunneling regime $w_{\rm in}\ll w$ at small frequencies,
$\omega\ll \Delta_{mn}$.
In contrast to sequential tunneling, where shot noise
is either Poissonian ($F=1$) or suppressed due to charge
conservation ($F<1$), we find that the noise in the
inelastic cotunneling regime can be super-Poissonian ($F>1$),
with a correction being as large as the Poissonian noise itself.
In the regime of elastic cotunneling $F=1$.

While the amount of super-Poissonian noise is merely estimated at the
end of Sec.~\ref{microscopic}, the noise of the cotunneling current
is calculated for the special case of a QDS with nearly degenerate states,
i.e.\ $\Delta_{nm}\ll \delta E$, in Sec.~\ref{degenerate}, where we
apply our results from  Sec.~\ref{microscopic}. The general solution
Eq.~(\ref{result01}) is further analyzed for {\em two} nearly degenerate
levels, with the result Eq.~(\ref{two-noise}).  More
information is gained in the specific case of a DD coupled to leads,
where we determine the correction to noise Eq.~(\ref{DD-noise})
as a function of frequency, bias, and the Aharonov-Bohm phase threading
the tunneling loop, finding signatures of the Aharonov-Bohm effect in
the cotunneling noise.

Finally, in Sec.~\ref{continuum}, another important situation is studied
in detail, the cotunneling through a QDS with a continuous energy spectrum,
$\delta E\ll\Delta\mu\ll E_C$.  Here, the correlation between tunneling
events plays a minor role as a source of super-Poissonian noise,
which is now caused by heating effects opening the possibility for
tunneling events in the reverse direction and thus to an enhanced
noise power.  In Eq.~(\ref{F-continuum}), we express the Fano factor $F$
in the continuum case in terms of the dimensionless numbers $C_\pm$,
defined in Eq.~(\ref{Cpm}), which depend on the electronic 
distribution function $f(\varepsilon)$ in the QDS (in this regime, a
description on the single-electron level is appropriate).
The current Eq.~(\ref{I-continuum}) is expressed in terms 
of the prefactor $\Lambda$, Eq.~(\ref{prefactor}).
Both $F$ and $\Lambda$ are then calculated for different regimes.
For weak cotunneling,
we immediately find $F=1$, as anticipated earlier, while for 
strong cotunneling we distinguish the two regimes of
cold ($\tau_{ee}\gg\tau_c$) and hot ($\tau_{ee}\ll\tau_c$) electrons.
For cold electrons, we derive the linear integral equation
Eq.~(\ref{kineq2}) for $f(\varepsilon)$ which is shown to have a
unique solution, and which is solved numerically.
We find that the Fano factor is very close
to one, $1<F<1.006$, while $\Lambda$ is given in Fig.~\ref{cold-hot}.
For hot electrons,  $f(\varepsilon)$ is the equilibrium
Fermi distribution, and the Fano factor Eq.~(\ref{F-hot}) and
$\Lambda$ [Eq.~(\ref{I-hot})and Fig.~\ref{cold-hot}] can be
computed analytically. Again, the Fano factor is very close to one,
$1<F<1.007$, which leads us to the conclusion that heating will
hardly be observed in noise, but should be well measurable
in the cotunneling current.

\acknowledgements
We are grateful to H. Schoeller for useful comments.
This work has been partially supported by the Swiss National
Science Foundation.

\appendix

\section{}
\label{A}

In this Appendix we present the derivation of Eqs.~(\ref{DB-current})
and (\ref{DB-noise}).
First we would like to mention that the operator $B$ in these equations
is just the second-order tunneling amplitude, which also appears
in the tunneling Hamiltonian after the Schrieffer-Wolff transformation.
Therefore, one might think that the Schrieffer-Wolff transformation 
is the most simple way to derive Eqs.~(\ref{DB-current}) and (\ref{DB-noise}).
On the other hand, it is obvious that the Schrieffer-Wolff procedure being a unitary
transformation gives exactly the same amount of terms in the fourth-order expression
for the current and noise as that of the regular perturbation expansion.
The Schrieffer-Wolff procedure is useful in the Kondo regime where the energy scale 
is given by the Kondo temperature $T_K$ and where the
$B$-terms in the Hamiltonian lead to a divergence for $T<T_K$, 
while the other terms can be treated by perturbation theory
(see Ref.~\onlinecite{Schrieffer}). 
In our cotunneling regime such a divergence does not
exist (since the QDS is weakly coupled to leads, i.e.\ 
$\Delta\mu,k_BT\gg k_BT_K$), 
and we have to analyze all contributions.
We do this below using perturbation theory.

In order to simplify the intermediate steps, we use the
notation $\bar O(t)\equiv\int_{-\infty}^{t}dt'O(t')$ for any operator $O$,
and $O(0)\equiv O$.
We notice that, if an operator $O$ is a linear function of operators $D_l$ and $D_l^{\dag}$,
then $\bar O(\infty)=0$ (see the discussion in Sec. \ref{FDT-Double}).
Next, the currents can be represented as the difference and the sum of $\hat I_1$ 
and $\hat I_2$,
\begin{eqnarray}
\hat I_d &=& (\hat I_2-\hat I_1)/2=ie(X^{\dag}-X)/2\,,
\label{A01}\\
\hat I_s &=& (\hat I_1+\hat I_2)/2=ie(Y^{\dag}-Y)/2\,,
\label{A02}
\end{eqnarray}
where $X=D_2+D_1^{\dag}$, and $Y=D_1+D_2$. 
While for the perturbation we have
\begin{equation}
V=X+X^{\dag}=Y+Y^{\dag}\,.
\label{A03}
\end{equation}
First we concentrate on the derivation
of Eq.~(\ref{DB-current}) and redefine the average current Eq.~(\ref{current-noise}) as
$I=I_d$ (which gives the same result anyway, because the average number
of electrons on the QDS does not change $I_s =0$).

To proceed with our derivation, we make use of Eq.~(\ref{U-Operator}) and expand
the current up to fourth order in $T_{lkp}$:
\begin{eqnarray}
I &=& i\int\limits^{0}_{-\infty}dt\int\limits^{t}_{-\infty}dt'
\langle \hat I_dV(t)V(t')\bar V(t')\rangle \nonumber \\
&-& i\int\limits^{0}_{-\infty}dt \langle \bar V\hat I_dV(t)\bar V(t)\rangle
+ {\rm c.c.}\,
\label{A04}
\end{eqnarray}
Next, we use the cyclic property of trace to shift the time dependence to $\hat I_d$.
Then we complete the integral over time $t$ and use $\bar I_d(\infty)=0$. This procedure
allows us to combine first and second term in Eq.~(\ref{A04}),
\begin{equation}
I=-i\int\limits^{0}_{-\infty}dt
\langle [\bar I_dV+\bar V\hat I_d]V(t)\bar V(t)\rangle+{\rm c.c.}
\label{A05}\,
\end{equation}
Now, using Eqs.~(\ref{A01}) and (\ref{A03})
we replace operators in Eq.~(\ref{A05}) with $X$ and $X^{\dag}$ in two steps:
$I=e\int^{0}_{-\infty}dt
\langle [\bar X^{\dag}X^{\dag}-\bar XX]V(t)\bar V(t)\rangle+{\rm h.c.}$,
where some terms cancel exactly. Then we work with $V(t)\bar V(t)$ and
notice that some terms cancel, because they are linear in $c_{lk}$ and
$c_{lk}^{\dag}$. Thus we obtain
\begin{eqnarray}
I = e\int\limits^{0}_{-\infty}dt
&&
\langle [\bar X^{\dag}X^{\dag}-\bar XX]\nonumber \\
&&\times [X^{\dag}(t)\bar X^{\dag}(t)+X(t)\bar X(t)]\rangle
+{\rm c.c.}\,
\label{A06}
\end{eqnarray}
Two terms $\bar XXX\bar X$ and
$\bar X^{\dag}X^{\dag}X^{\dag}\bar X^{\dag}$ describe tunneling
of two electrons from the same lead, and therefore they do not contribute
to the normal current. We then combine all other terms to extend the integral to $+\infty$,
\begin{equation}
I = e\int\limits^{\infty}_{-\infty}dt
\langle\bar X^{\dag}(t)X^{\dag}(t)X\bar X
-\bar XXX^{\dag}(t)\bar X^{\dag}(t)\rangle\,
\label{A07}
\end{equation}
Finally, we use
$\int^{\infty}_{-\infty}dt X(t)\bar X(t)=-\int^{\infty}_{-\infty}dt \bar X(t)X(t)$
(since $\bar X(\infty)=0$) to get Eq.~(\ref{DB-current}) with $B=X\bar X$.
Here, again, we drop
terms $D^{\dag}_1\bar D^{\dag}_1$ and $D_2\bar D_2$ responsible for
tunneling of two electrons from the same lead, and obtain $B$
as in Eq.~(\ref{DB-amplitude}).

Next, we derive Eq.~(\ref{DB-noise}) for the noise power.
At small frequencies $\omega\ll \Delta_{\pm}$ fluctuations
of $I_s$ are suppressed because of charge conservation (see below),
and we can replace $\hat I_2$ in the correlator Eq.~(\ref{current-noise}) with $\hat I_d$.
We expand $S({\omega})$ up to fourth order in $T_{lkp}$,
use $\int_{-\infty}^{+\infty}dt\, \hat I_d(t)e^{\pm i\omega t}=0$,
and repeat the steps leading to Eq.~(\ref{A05}). Doing this we
obtain,
\begin{equation}
S(\omega)=-\int\limits^{\infty}_{-\infty}dt \cos(\omega t)
\langle [\bar V(t),\hat I_d(t)][\bar V, \hat I_d]\rangle\,.
\label{A08}
\end{equation}
Then, we replace $V$ and $\hat I_d$ with $X$ and $X^{\dag}$.
We again keep only terms relevant
for cotunneling, 
and in addition we neglect terms of order $\omega/\Delta_{\pm}$
(applying same arguments as before, see Eq.~(\ref{A09})).
We then arrive at Eq.~(\ref{DB-noise})
with the operator $B$ given by Eq.~(\ref{DB-amplitude}).

Finally, in order to show that fluctuations of $I_s$ are suppressed,
we replace $\hat I_d$ in Eq.~(\ref{A08}) with $\hat I_s$, 
and then use the operators
$Y$ and $Y^{\dag}$ instead of $X$ and $X^{\dag}$. 
In contrast to Eq.\ ({\ref{A07}}) terms such as 
$\bar Y^{\dag}Y^{\dag}Y\bar Y$ do not contribute,
because they contain integrals  of the form
$\int^{\infty}_{-\infty}dt\cos(\omega t) D_{l}(t)\bar D_{l'}(t)=0$.
The only nonzero contribution can be written as
\begin{equation}
S_{ss}(\omega)=\frac{e^2\omega^2}{4}\int\limits^{\infty}_{-\infty}dt \cos(\omega t)
\langle [\bar Y^{\dag}(t),\bar Y(t)][\bar Y^{\dag},\bar Y]\rangle\,,
\label{A09}
\end{equation}
where we have used integration by parts and the property $\bar Y(\infty)=0$.
Compared to Eq.~(\ref{DB-noise}) this expression contains an additional
integration over $t$, and thereby it is of order
$(\omega/\Delta_{\pm})^2$.

\section{}
\label{B}
We evaluate the matrix elements of the superoperator $W^I(z)$
given in Eq.~(\ref{expval-2}) which are used to calculate the
average current $I_l$, see Eq.~(\ref{AvCurrent}).
The derivation for the master equation (\ref{MasterEq})
is very similar.
As for the noise, the $S^Q_{ll'}$ term Eq.~(\ref{correlator-q})
is again obtained in a similar way as the current, whereas
the $S^P_{ll'}$ term Eq.~(\ref{correlator-p-1})
is different and is analyzed in Sec.~\ref{microscopic_correlators}.
Since $W^I(z)$ is obtained by
taking the partial trace over the leads, its matrix elements
can be expressed as the sum over lead indices
\begin{equation}
  \label{B1}
  W^I_{nm}(z) =
  \sum_{\bar{n}\bar{m}} {\cal W}^I_{\bf nm}(z) \, \rho_{L, \bar{m}} (z),
\end{equation}
where ${\bf n}=(n,\bar{n})$,
with $n$ and $\bar{n}$ enumerating the QDS and lead eigenstates.
For convenience, we will use the eigenstates of $H_0$ in this
Appendix, and not the eigenstates of $K$ as in the main text.
Accordingly, here $E_{\bf n}=E_n+E_{\bar{n}}$ are the eigenenergies
of $H_0$.
Taking the stationary limit $z \rightarrow 0$, using the definition
Eq.~(\ref{expval-2}) and
introducing the projectors $p_{\bf n}=|{\bf n}\rangle \langle {\bf n}|$,
we can write
\begin{equation}
  \label{B2}
  {\cal W}^I_{\bf nm} = \lim_{z\rightarrow 0}
  {\rm Tr}\, p_{\bf n} \hat I_l Q \frac{1}{z-QLQ}QL_VPp_{\bf m}.
\end{equation}
Note that while ${\bf n}$ denotes a free dummy index in Eq.~(\ref{B2}),
the state $|{\bf m}\rangle$ is restricted to the subspace
where $P_N p_{\bf m}\neq 0$ with fixed particle
number $N$ on the QDS.
Expanding this expression in $V$, we obtain for the lowest nonvanishing order
(sequential tunneling) the contribution
$-i\sum_{\bar{n}\bar{m}}(\hat I_l R_0 L_V p_{\bf m})_{\bf nn}$
to the rate $W^I_{nm}$, which can be expressed as
\begin{equation}
  \label{B3}
  2\pi e\sum_{\bar{n}\bar{m}} \left(|\langle {\bf n}|D_l|{\bf m}\rangle|^2
   -|\langle{\bf n}|D_l^\dagger|{\bf m}|\rangle|^2\right)
       \rho_{L,\bar{m}}\,\delta(\Delta_{\bf mn}),
\end{equation}
where $\Delta_{\bf mn}=E_{\bf m}-E_{\bf n}$.  Using Eq.~(\ref{tunneling})
and assuming that $T_{lkp}={\cal T}$ is independent of $p$ and $k$,
we obtain the
expression for the contribution to $W^I_{nm}$ due to sequential tunneling,
\begin{eqnarray}
2\pi\nu {\cal T}^2 \sum_p
 \left(|\langle n|d_p|m\rangle |^2 \left[1-f_l(\Delta_{mn})\right]\right.&&
      \nonumber\\
      \quad\quad\quad
      \left.-|\langle n|d_p^\dagger |m\rangle|^2 f_l(\Delta_{nm})\right),&&
  \label{B4}
\end{eqnarray}
where $f_l(\varepsilon)$ is the Fermi distribution and $\nu$ the density
of states in the leads.
In the cotunneling regime\cite{cotunneling1},
this contribution is proportional to $\kappa=e^{-\Delta/k_BT}$,
therefore we drop it\cite{cotunneling2} and
expand $W^I_{nm}$ to the next non-vanishing, i.e.\ fourth,
order in $V$. Doing this, we obtain the cotunneling
contribution
\begin{equation}
  {\cal W}^I_{\bf nm} 
    = i(\hat I_l R_0 L_V R_0 Q L_V R_0 L_V p_{\bf m})_{\bf nn}.
  \label{B5}
\end{equation}
Stepwise evaluation of the operators and superoperators in this
expression by the insertion of the
identity $\sum_{\bf i}|{\bf i}\rangle\langle {\bf i}|$ leads to
\begin{eqnarray}
  {\cal W}^I_{\bf nm}
      &=& i \sum_{{\bf i},{\bf j}}
           ( I_{\bf ni} R_{\bf in} V_{\bf ij} R_{\bf jn} U_{\bf jn}^{\bf m}
           - I_{\bf ni} R_{\bf in} R_{\bf ij} U_{\bf ij}^{\bf m} V_{\bf jn} ),
\nonumber\\
U_{\bf ij}^{\bf m} &=& (L_V R_0 L_V p_{\bf m})_{\bf ij}\nonumber\\
                       &=& \sum_{\bf k} \left[ V_{\bf ik}R_{\bf kj}(L_Vp_{\bf m})_{\bf kj}
                        -   R_{\bf ik}(L_Vp_{\bf m})_{\bf ik} V_{\bf kj}\right],
\nonumber\\
  (L_Vp_{\bf m})_{\bf ij}
    &=& V_{\bf im}\delta_{\bf mj} - V_{\bf mj}\delta_{\bf im},
\label{B6}
\end{eqnarray}
where $I_{\bf ij}=\langle {\bf i}|\hat I_l|{\bf j}\rangle$, and
similarly for $V_{\bf ij}$.  Note that
\begin{eqnarray}
  R_{\bf ij} &=& \lim_{\eta\rightarrow 0}
                      \frac{i}{i\eta-(E_{\bf i}-E_{\bf j})}
\nonumber\\
             &=& -i {\rm P}\frac{1}{E_{\bf i}-E_{\bf j}}
                 +\pi\delta(E_{\bf i}-E_{\bf j}),
\label{B7}
\end{eqnarray}
where P stands for the principal value.
The current $I_l$ is obtained from ${\cal W}^I_{\bf nm}$ by multiplying
with the full density matrix $\rho_{\bf m}$ and then 
summing over ${\bf m}$ and ${\bf n}$.
By explicit evaluation, using the fact that we can choose the basis $|n\rangle$ on the
QDS such that all expectation values of the form
$\langle n|d_{p_1}^\dagger d_{p_2} d_{p_3}^\dagger  d_{p_4}|n\rangle$, etc., are real,
we find that four out of the eight terms in Eq.~(\ref{B6}) cancel, while the
remaining four terms contributing to the current $I_l$ can be combined into
(retaining only $O(\kappa^0)$ terms)
\begin{eqnarray}
\sum_{\bf n} {\cal W}^I_{\bf nm} &=& -2\pi \,{\rm Im}\sum_{\bf f}\left[
(\hat I_lR_{\bf m}^\dagger V)_{\bf mf} (VR_{\bf m}^\dagger V)_{\bf fm}\right.
\nonumber\\
&&\left.+ (VR_{\bf m}V)_{\bf mf} (\hat I_l R_{\bf f}^\dagger V)_{\bf fm}
\right]\delta(E_{\bf f}-E_{\bf n}),
\label{B8}
\end{eqnarray}
where $R_{\bf m}=-i{\rm P}(H_0-E_{\bf m})^{-1}$.
All other $\delta$-function contributions vanish in $O(\kappa^0)$.
\cite{cotunneling2}
In the presence of an Aharonov-Bohm phase, when the phases in the
tunneling amplitudes Eq.~(\ref{equal}) have to be taken into account, we
again find Eq.~(\ref{B8}) by explicit analysis.
We note here that exactly the same procedure as above can be applied in the
derivation of the the master equation and the noise, leading to a reduction
of terms and finally to the ``golden rule'' expressions Eqs.~(\ref{master-6})
and (\ref{correlator-q}).
By substituting Eqs.~(\ref{tunneling}) and (\ref{currents})
for $V$ and $\hat I_l$, and setting $l=2$ for concreteness, we finally obtain
\begin{eqnarray}
  \sum_{\bf n} {\cal W}^I_{\bf nm}
  &=&  2\pi e\sum_{\bf f} \left[ (D_2^\dagger, D_1)_{\bf mf}(D_1^\dagger, D_2)_{\bf fm}
\right.
\nonumber\\
&& \left.- (D_1^\dagger, D_2)_{\bf mf}(D_2^\dagger, D_1)_{\bf fm}\right]
   \delta(\Delta_{\bf fm}),
  \label{B9}
\end{eqnarray}
where $(D_l^\dagger , D_{l'})$ is defined in Eq.~(\ref{matrix-element}).
Using Eqs.~(\ref{expval-2}) and (\ref{B1}) and the definitions Eqs.~(\ref{w-pm})
and (\ref{golden-rule}),
we find for the cotunneling current
\begin{equation}
  \label{B10}
  I_2 = \sum_{\bf nm}{\cal W}^I_{\bf nm}\,\bar\rho_{m}\rho_{L,\bar{m}}
      = e\sum_{nm} (w^+_{nm} - w^-_{mn})\bar\rho_{m},
\end{equation}
which concludes the derivation of Eqs.~(\ref{AvCurrent}) and (\ref{w-current}).
Note that in Eq.~(\ref{golden-rule}) the expression
$\Delta_{\bf mn}=E_{\bf m}-E_{\bf n}$ is replaced
by $E_{\bf m}-E_{\bf n}-\Delta\mu_{ll'}$  because
there, $|{\bf n}\rangle$ are eigenstates of $K$ (instead of $H_0$).
The current $I_1$ in lead $1$ 
can be obtained by interchanging the lead indices $1$ and $2$
in Eq.~(\ref{B9}) which obviously leads to $I_1=-I_2$.

\section{}
\label{C}
In this Appendix we calculate the transition rates Eq.~(\ref{golden-rule}) for 
a DD coupled to leads with the coupling described by Eqs.~(\ref{DD-D})
and (\ref{equal}) and show that the four-level system in the singlet-triplet basis
Eq.~(\ref{basis})
can be mapped to a two-level system. For the moment we assume that the indices
$n$ and $m$ enumerate
the singlet-triplet basis, $n,m = S,T_0,T_{+},T_{-}$. Close to the sequential tunneling
peak, $\Delta_{-}\ll\Delta_{+}$, we keep only terms of the form $D^{\dag}_{l}R_0D_{l}$.
Calculating the trace over the leads explicitly, we obtain at $T=0$,
\begin{eqnarray}
w_{nm}(l',l)
&=&
\frac{\pi\nu^2}{2\Delta^2_{-}}\,
\Theta(\mu_l-\mu_{l'}-\Delta_{nm})
\nonumber \\
& \times &
\sum_{j,j'}T^{*}_{lj}T_{lj'}T^{*}_{l'j'}T_{l'j}
M_{nm}(j,j')\,,
\label{C1}\\
M_{nm}(j,j')
&=& \sum_{s,s'}\langle n |d^{\dag}_{sj}d_{s'j}|m\rangle
\langle m |d^{\dag}_{s'j'}d_{sj'}|n\rangle\,,
\label{C2}
\end{eqnarray}
with 
$\Theta(\varepsilon)=\varepsilon \theta(\varepsilon)$,
and $\Delta_{nm}=0, \pm J$, and we have assumed $t_d\ll \Delta_{-}$
so that $R_0=1/\Delta_{-}$.

Since the quantum dots are the same we get $M_{nm}(1,1)=M_{nm}(2,2)$,
and $M_{nm}(1,2)=M_{nm}(2,1)$. 
We calculate these matrix elements in the singlet-triplet basis
explicitly,
\begin{eqnarray}
&& M(1,1)=\frac{1}{2}\left(
\begin{array}{rrrr}
 1&1&1&1\\
1&1&1&1\\
1&1&2&0\\
1&1&0&2
\end{array}
\right),  
\label{C3} \\
&&M(1,2)=\frac{1}{2}\left(
\begin{array}{rrrr}
1&-1&-1&-1\\
-1&1&1&1\\
-1&1&2&0\\
-1&1&0&2
\end{array}
\right).  
\label{C4}
\end{eqnarray}
Assuming now equal coupling of the form Eq.~(\ref{equal})
we find that for $l=l'$ the matrix elements of the singlet-triplet
transition vanish (as we have expected, see Sec.\ \ref{degenerate}).
On the other hand the triplets are degenerate, i.e.\ $\Delta_{nm}=0$
in the triplet sector. Then from Eq.~(\ref{C1}) it follows
that $w^{0}_{nm}=\sum_{l}w_{nm}(l,l)=0$.
Next, we have $\Theta(\mu_2-\mu_{1}-\Delta_{nm})=0$,
since for nearly degenerate states we assume $\Delta\mu>|\Delta_{nm}|$,
and thus $w^{-}_{nm}=w_{nm}(1,2)=0$.
Finally, for $w^{+}_{nm}=w_{nm}(2,1)$ we obtain,
\begin{eqnarray}
w^{+}_{SS}
&=&
\frac{\pi}{2}\left(\frac{\nu{\cal T}^2}{\Delta_{-}}\right)^2     
\Delta\mu(1+\cos\phi),
\label{C5-1} \\
w^{+}_{ST}
&=&
\frac{\pi}{2}\left(\frac{\nu{\cal T}^2}{\Delta_{-}}\right)^2     
(\Delta\mu+J)(1-\cos\phi),
\label{C5-2}\\
w^{+}_{TS}
&=&
\frac{\pi}{2}\left(\frac{\nu{\cal T}^2}{\Delta_{-}}\right)^2     
(\Delta\mu-J)(1-\cos\phi),
\label{C5-3}\\
w^{+}_{TT}
&=&
\frac{\pi}{2}\left(\frac{\nu{\cal T}^2}{\Delta_{-}}\right)^2     
\Delta\mu
\nonumber\\
&\times&
\left(
\begin{array}{lll}
1+\cos\phi & 1+\cos\phi & 1+\cos\phi \\
1+\cos\phi & 2+2\cos\phi & 0\\ 
1+\cos\phi & 0 & 2+2\cos\phi  
\end{array}
\right).
\label{C5-4}
\end{eqnarray}

Next we prove the mapping to a two-level system. 
First we notice that because the matrix $w^{+}_{TT}$ is symmetric, 
the detailed balance equation for the stationary state gives 
$\bar\rho_n/\bar\rho_m=w^{+}_{mn}/w^{+}_{nm}=1$, $n,m\in T$.
Thus we can set $\bar\rho_{n}\to\bar\rho_2/3$, for $n\in T$.
The specific form of the transition matrix Eqs.~(\ref{C5-1}-\ref{C5-4})
helps us to complete the mapping by setting 
$(1/3)\sum_{m=2}^4w^{+}_{1m}\to w^{+}_{12}$,
$\sum_{n=2}^4w^{+}_{n1}\to w^{+}_{21}$, and
$(1/3)\sum_{n,m=2}^4w^{+}_{nm}\to w^{+}_{22}$, so that 
we get the new transition matrix Eq.~(\ref{DD-rates}), while 
the stationary master equation for the new two-level density matrix
does not change its form. If in addition we set 
$(1/3)\sum^{4}_{m=2}\delta\rho_{1m}(t)\to\delta\rho_{12}(t)$,
$\sum^4_{n=2}\delta\rho_{n1}(t)\to\delta\rho_{21}(t)$, and
$(1/3)\sum^{4}_{n,m=2}\delta\rho_{nm}(t)\to\delta\rho_{22}(t)$,
then the master equation Eq.~(\ref{MasterEq}) for $\delta\rho_{nm}(t)$ and
the initial condition $\delta\rho_{nm}(0)=\delta_{nm}-\bar\rho_{n}$
do not change either. Finally, one can see that under this mapping
Eq.~(\ref{correlator-delta-s}) 
for the correction to the noise power $\Delta S(\omega)$
remains unchanged. Thus we have accomplished 
the mapping of our singlet-triplet system 
to the two-level system with the new transition 
matrix given by Eq.~(\ref{DD-rates}). 

\section{}
\label{D}

Here we prove that the solution of Eqs.~(\ref{kineq2}),
(\ref{Upsilon}), and (\ref{conditions}) is not degenerate. Suppose
the opposite is true, i.e.\ there are two functions,
$f_1(\varepsilon)$ and $f_2(\varepsilon)$, which satisfy these
equations. Then the function
$f_d(\varepsilon)=f_1(\varepsilon)-f_2(\varepsilon)$ satisfies 
Eq.~(\ref{kineq2}) with the conditions
\begin{eqnarray}
&& \int d\varepsilon f_d(\varepsilon)=\int d\varepsilon \varepsilon f_d(\varepsilon)=0,
\label{condition1} \\
&& f_d(+\infty)=f_d(-\infty)=0,\quad -1\leq f_d(\varepsilon)\leq 1.
\label{condition2}
\end{eqnarray}
According to Eqs.~(\ref{kineq2}), and (\ref{Upsilon}),
the integral $\int d\varepsilon |\varepsilon f_d(\varepsilon)|$ is convergent.
This allows us to symmetrize the kernel $\sigma$ in Eq.~(\ref{kineq2}):
$\sigma(\varepsilon)=\sigma_S(\varepsilon)+(1+\lambda)\varepsilon +\Delta\mu$, where
$\sigma_S(\varepsilon)=[\lambda\Theta(\varepsilon)
+\Theta(\varepsilon-\Delta\mu)]+[\varepsilon\to -\varepsilon]$,
and thus
$\sigma_S(\varepsilon)=\sigma_S(-\varepsilon)$.
Using the condition Eq.~(\ref{condition1}) we arrive at the new integral
equation for $f_d$,
\begin{eqnarray}
&& \int d\varepsilon'\sigma_S(\varepsilon'-\varepsilon)f_d(\varepsilon')
\nonumber \\
&&\qquad\qquad\qquad
=[(1+\lambda)(\varepsilon^2+2\Upsilon)+(\Delta\mu)^2]f_d(\varepsilon).
\label{kineq5}
\end{eqnarray}

Next we apply Fourier transformation to both sides of this equation
and introduce the function
\begin{equation}
\varphi(x)=\frac{1}{2\pi}\int d\varepsilon e^{-i\varepsilon x}f_d(\varepsilon).
\label{varphi}
\end{equation}
Here we have to be careful because, strictly speaking the
Fourier transform of $\sigma_S(\varepsilon)$ does not exist
(this function is divergent at $\pm\infty$). On the other hand,
since the integral on the lhs of Eq.~(\ref{kineq5}) is 
convergent, we can regularize the kernel as
$\sigma_S(\varepsilon)\to\sigma_S(\varepsilon)e^{-\eta|\varepsilon|}$
and later take the limit $\eta\to +0$. Then for the Fourier
transform of Eq.~(\ref{kineq5}) we find
\begin{eqnarray}
&& (1+\lambda)\varphi''(x)=[u(x)+(\Delta\mu)^2+2(1+\lambda)\Upsilon]\varphi(x),
\label{diffeq} \\ &&
u(x)=\int d\varepsilon e^{-i\varepsilon x}\sigma_S(\varepsilon)
= 2[\lambda+\cos(\Delta\mu x)]/x^2,
\label{potential}
\end{eqnarray}
where $u(x)$ is real, because $\sigma_S$ is an even function of
$\varepsilon$. Thus we have obtained a second order differential
(Schr\"odinger) equation for the function $\varphi(x)$.
We conclude from Eq.~(\ref{condition1}) that
$\varphi(0)=\varphi'(0)=0$, and the condition Eq.~(\ref{condition2})
ensures that the solution of Eq.~(\ref{diffeq}) is localized,
$\varphi(x)|_{x\to\pm\infty}=0$ and finite everywhere. All
these requirements can be satisfied only if $\varphi(x)=0$ for all
$x$. Indeed, since the function $u(x)+(\Delta\mu)^2+2(1+\lambda)\Upsilon$
is positive for all $x$ (we recall that $\Upsilon >0$), then
$\varphi$ is a monotonous function, and therefore it cannot be
localized. In other words, the Schr\"odinger equation with repulsive
potential $u(x)>0$ does not have localized solutions. Thus we have proven
that $f_1(\varepsilon)=f_2(\varepsilon)$ for all $\varepsilon$,
and the solution of Eq.~(\ref{kineq2}) is not degenerate.


\begin{references}
\bibitem{comment1}
If the displacement current is taken into account,
then the situation becomes more complicated,
\cite{Halperin,Hanke1,Hanke2}
while our general physical picture is still valid.
In addition, we note that for cotunneling
the displacement current can be neglected because
there is no charge accumulation on the QDS.
\bibitem{Josephson-frequency}
The other quantum frequency scale is given
by the Josephson frequency $E_J$ as, for example,
in the S-S-N junction\cite{Mahn-Soo}.
\bibitem{cotunneling1}
We formally define the cotunneling regime 
through the condition $\kappa=\exp(-\Delta/k_BT)\ll 1$, where $T$ is the temperature
and $\Delta = \min\limits_{\pm, l}\{\Delta_\pm(N,l)\}$ is the minimum energy which is required
to transfer an electron between the leads and the QDS.
Physically, this means that we are sufficiently far away from the sequential
tunneling resonance to neglect fluctuations of the particle number on the QDS (see also Sec.~\ref{me}).
\bibitem{comment3}
In case of disordered QDS with the Thouless energy $E_{\rm th}<E_C$
the condition for the low bias regime has to be replaced by
$\Delta\mu\ll (\delta E\, E_{\rm th})^{1/2}$,
see Ref.\ \onlinecite{averinazarov}.
\bibitem{comment4}
We note that charge fluctuations,
$\delta Q(t)\!=\!2\!\int_{-\infty}^{t}\!dt' \delta I_s(t')$,
on a QDS are also relevant for device
applications such as single-electron transistors,
see Ref.~\onlinecite{problem1}.
While we focus on current fluctuations in the present paper, we mention here
that in the cotunneling regime
the noise power $\langle \delta Q^2\rangle_{\omega}$
does not vanish at zero frequency,
$\langle \delta Q^2\rangle_{\omega=0}=
4\omega^{-2}S_{ss}(\omega)|_{\omega\to 0}\neq 0$.
Our formalism is also suitable for studying such charge fluctuations;
this will be addressed elsewhere.
\bibitem{comment5}
One could view this as an analog of a whistle effect,
where the flow of air (current) is strongly modulated by a bistable state
in the whistle, and vice versa. The analogy, however, is not
complete,
since the current through the QDS is random due to quantum fluctuations.
\end{references}
\end{document}